\documentclass[amsmath,amssymb,nofootinbib,superscriptaddress,showpacs,showkeys]{revtex4-2}
\usepackage{graphicx}
\usepackage{dcolumn}
\usepackage{bm}
\usepackage{rotating} 
\usepackage{subfigure}
\usepackage{booktabs}
\usepackage{multirow}
\usepackage{siunitx}
\usepackage{physics}
\usepackage{multirow} 
\usepackage[table]{xcolor}
\usepackage[colorlinks=true, citecolor=blue, urlcolor = magenta, linkcolor= red, bookmarks=true]{hyperref}
\usepackage{orcidlink}

\begin{document}

\title{Schwarzschild-AdS Black Holes with Cloud of Strings and Quintessence: Geodesics, Thermodynamic Topology, and Quasinormal Modes }

\author{Faizuddin Ahmed\orcidlink{0000-0003-2196-9622}} \email{faizuddinahmed15@gmail.com} 
\affiliation{Department of Physics, Royal Global University, Guwahati, 781035, Assam, India}

\author{Saeed Noori Gashti\orcidlink{0000-0001-7844-2640}} \email{saeed.noorigashti70@gmail.com; sn.gashti@du.ac.ir} 
\affiliation{School of Physics, Damghan University, P. O. Box 3671641167, Damghan, Iran}

\author{Abdelmalek Bouzenada\orcidlink{0000-0002-3363-980X}}\email{ abdelmalekbouzenada@gmail.com}
\affiliation{Laboratory of Theoretical and Applied Physics, Echahid Cheikh Larbi Tebessi University 12001, Algeria}

\author{Behnam Pourhassan\orcidlink{0000-0003-1338-7083}} \email{b.pourhassan@du.ac.ir} 
\affiliation{School of Physics, Damghan University, P. O. Box 3671641167, Damghan, Iran}
\affiliation{Center for Theoretical Physics, Khazar University, 41 Mehseti Street, Baku, AZ1096, Azerbaijan}

\begin{abstract}
In this study, we explore a Schwarzschild-anti de-Sitter black hole (BH) coupled with a cloud of strings (CoS) possessing both electric- and magnetic-like components of the string bivector, embedded in a Kiselev-type quintessence fluid (QF). We analyze the dynamics of photons and test particles, focusing on trajectories, photon spheres, BH shadows, and innermost stable circular orbits (ISCO), highlighting how CoS and QF parameters affect these features. We then examine the thermodynamic topology of the system by analyzing vector field zeros, showing that varying CoS leads to distinct topological configurations with total charges of either $0$ or $+1$, corresponding to known classes like RN and AdS-RN. Additionally, we study scalar field dynamics via the massless Klein-Gordon equation, reformulated into a Schrödinger-like form to derive the effective potential. We compute the quasinormal modes (QNMs) of scalar perturbations, showing how CoS and QF influence oscillation frequencies and damping rates, with implications for gravitational confinement and thermalization in the AdS/CFT context. 
\end{abstract}
\maketitle

\section{Introduction} \label{S1}

The detection of gravitational waves (GWs) by the LIGO-Virgo collaboration~\cite{BH1,BH2}, along with the first image of a supermassive BH (BH) in M87 by the Event Horizon Telescope (EHT)~\cite{BH3,BH4,BH5,BH6,BH7,BH8}, has significantly advanced both observational and theoretical relativistic astrophysics. These milestones not only confirm key predictions of general relativity (GR) but also provide opportunities to test gravity in the strong-field regime. While GR continues to pass weak-field tests, such as in the solar system~\cite{BH9}, current precision limits from GW and BH shadow observations still allow for potential deviations~\cite{BH10,BH11}, motivating investigations into how alternative gravity parameters influence observable BH features.

Recent work has extended BH shadow analysis using wave optics, particularly through the approach proposed by Hashimoto et al.~\cite{SH1}, showing structures such as Einstein rings in Schwarzschild-AdS backgrounds~\cite{SH2,SH3}. The role of charge has also been explored~\cite{SH4}, demonstrating its impact on both optical and thermodynamic properties. Among BH models, the Schwarzschild--AdS solution remains central due to its relevance in gravitational theory and the AdS/CFT correspondence~\cite{SH5,SH6,SH7}. Modifications to this geometry, including those via gravitational decoupling techniques~\cite{SH8,SH9}, affect event horizon radii, phase transitions, and thermal stability~\cite{SH10}. Unlike their flat-space counterparts, AdS BHs exhibit positive specific heat, enabling thermal equilibrium at high energies~\cite{SH11}, consistent with CFT behavior at the boundary~\cite{SH12}.

Quasinormal modes (QNMs) are the characteristic damped oscillations emitted by a BH when perturbed-akin to the "ringing" of a struck bell but for space-time itself. These modes are defined by complex frequencies, where the real part represents the oscillation frequency and the imaginary part indicates the decay rate, reflecting how rapidly the perturbation fades over time~\cite{QNMs1,QNMs2,QNMs3,QNMs4}. 

In the context of BH stability and gravitational wave astronomy, QNMs play a vital role: they encode intrinsic properties such as mass, spin, and charge of the remnant BH, and are insensitive to the perturbation details, making them powerful probes of gravity in the strong-field regime~\cite{Q1,Q2}. They arise naturally as solutions to non-Hermitian eigenvalue problems under physically motivated boundary conditions: purely ingoing waves at the horizon and purely outgoing waves at infinity~\cite{Q1}. Analytically and numerically, QNMs have been extensively studied across various gravitational theories, such as scalar-tensor models and Gauss-Bonnet gravity~\cite{Q2,Q3}, and in higher-dimensional setups. Computational approaches include WKB methods, continued fractions, asymptotic iteration, and time-domain integration techniques. Moreover, within the AdS/CFT correspondence, BH QNMs correspond to the relaxation timescales in the dual strongly coupled field theories, offering insights into hydrodynamic behavior and thermalization-bridging the gap between classical gravity and quantum field theory~\cite{Q1,Q3}.

Quasinormal modes (QNMs) play a central role in gravitational wave (GW) physics, particularly in the ringdown phase of BH mergers. When two BHs coalesce, the remnant BH emits GWs that rapidly settle into a superposition of QNMs-damped oscillations uniquely determined by the mass, spin, and charge of the BH~\cite{Q1,Q2}. These modes provide a powerful spectroscopic tool, analogous to atomic spectral lines, allowing precise measurements of BH parameters and offering a direct way to test the no-hair theorem of general relativity~\cite{QQ3}.

The first detection of GWs from a binary BH merger, GW150914~\cite{GW1}, confirmed the existence of QNMs predicted by general relativity. Ongoing observations by LIGO-Virgo-KAGRA are refining QNM measurements, enabling stringent tests of the Kerr hypothesis and constraining deviations from general relativity~\cite{QQ4}. Furthermore, QNMs are crucial in identifying exotic compact objects, probing quantum gravity effects, and exploring BH area quantization~\cite{Q3,Q6}.

Recent developments in analytical techniques have introduced new pathways for studying phase transitions and criticality in BH (BH) systems. Among these innovations, topological approaches have emerged as powerful tools, offering a global view of the thermodynamic structure \cite{a19,a20}. These approaches are rooted in Duan's topological current theory, which has been effectively adapted to explore BH thermodynamics. In particular, the work of S.W. Wei and collaborators has established two prominent topological frameworks for BH phase structure analysis, the so-called T-method and F-method \cite{a19,a20}. The T-method treats temperature as the primary thermodynamic variable, deliberately separating it from pressure, and introduces auxiliary functions, such as $1/\sin\theta$, to define a scalar potential. Within this context, the temperature is treated as a vector field on a two-dimensional manifold, and topological charges are used to pinpoint critical phenomena. Although this technique efficiently identifies phase transition points, it offers limited insight into the microstructure or continuity of the transition itself. In contrast, the F-method employs the Helmholtz free energy as a scalar function over the thermodynamic parameter space. This method conceptualizes BHs as topological defects and investigates the free energy's geometric features, specifically, its curvature and associated winding numbers, to probe phase behavior. The emergence of multiple extrema in the free energy within a finite range of horizon radii typically signals a first-order phase transition, marked by the coexistence of multiple BH phases. If the free energy varies smoothly without intermediate extrema, a second-order transition is indicated \cite{a19,a20,20a,22a,23a,24a,25a,26a,27a,33a,38a,38b,38c,40a,42a,44f,44g,44h,44i,44j,44k,44l,44m,44n,44o}.
While both the T and F frameworks examine phase transitions from distinct thermodynamic perspectives, one emphasizing temperature dynamics and the other focusing on free energy, they collectively offer a more comprehensive topological classification of BH states. In particular, the F-method also allows for the assessment of thermodynamic stability through the sign of the second derivative of the free energy: a positive value implies stability, while a negative one suggests instability.

Beyond thermodynamics, the topological framework is increasingly applied to the study of photon spheres, regions near BHs where massless particles like photons can be trapped in circular orbits due to strong gravitational effects. These photon spheres are critical for understanding gravitational lensing, shadow formation, and other observable signatures of BHs. Using topological techniques, researchers investigate both the dynamical stability and geometric structure of these photon orbits, deepening our understanding of light behavior in the vicinity of BHs \cite{45'm,45mmm,45m,47m,48m,49m,50m}.

In this study, we investigate a Schwarzschild-AdS BH solution coupled with a CoS, incorporating both electric- and magnetic-like components of the string bivector, and surrounded by a QF. We begin by analyzing the geodesic motion of test particles, both massless and massive, focusing on photon trajectories, the photon sphere, BH shadows, circular null orbits, and the ISCO. We demonstrate how geometric and physical parameters-governing the space-time curvature-significantly affect photon and particle dynamics, including shifts in the photon sphere radius, shadow size, and ISCO location. Subsequently, we explore the thermodynamic topology of the BH space-time, showing that the system admits distinct topological configurations with total charges of either $0$ or $+1$, depending on the values of CoS and QF parameters. These configurations correspond to known BH classes such as Reissner-Nordström (RN) and AdS-RN, highlighting a connection between space-time geometry and thermodynamic phase structure. Finally, we study the dynamics of spin-0 scalar field perturbations in this BH solution background. By deriving the effective perturbative potential, we examine how CoS and QF parameters influence the potential profile. Using this, we numerically compute the quasinormal mode (QNM) frequencies and show how the geometric and physical parameters modify both oscillation frequencies and damping rates. Overall, our results demonstrate that the combined effect of CoS and QF alters the space-time curvature and, consequently, impacts both dynamical and thermodynamic properties of the BH solution.

This paper is structured as follows: Section~\ref{sec:2} introduces the theoretical framework of the Schwarzschild--AdS BH with a cloud of strings and a quintessence-like fluid. Section~\ref{sec:3} investigates the geodesic structure of the BH space-time, including photon trajectories, BH shadows, ISCO, and geodesic topology. Section~\ref{sec:4} examines the thermodynamic topology of the BH system and discusses the resulting topological configurations. Section~\ref{sec:5} analyzes the dynamics of spin-0 scalar field perturbations and presents the corresponding quasinormal mode (QNM) spectra. Finally, Section~\ref{sec:6} summarizes the main results and provides concluding remarks.

\section{AdS BH solution coupled with CoS embedded with QF}\label{sec:2}

The Einstein-Hilbert action coupled to a cosmological constant and a matter source describing a CoS is given by \cite{PSL,YBZ}:
\begin{equation}\label{action}
S = \frac{1}{16\pi G}\int d^4x\sqrt{-g}\left[R - 2\Lambda\right] + S_M,
\end{equation}
where $R$ is the Ricci scalar curvature, $\Lambda$ is the negative cosmological constant related to the AdS radius $l$ through $\Lambda = -3/\ell^2_p$ \cite{SWH}, and $S_M$ represents the matter action. The matter content is described by the Nambu-Goto action, which effectively models a CoS as proposed in \cite{PSL}:
\begin{equation}\label{NambuGoto}
S_M = \mathcal{M}\,\int_\Sigma \sqrt{-\gamma}\,d\lambda^0 d\lambda^1 = \mathcal{M}\, \int_\Sigma \left[-\frac{1}{2}\Sigma_{\mu\nu}\,\Sigma^{\mu\nu}\right]^{1/2} d\lambda^0 d\lambda^1.
\end{equation}
Here, $\lambda^a = (\lambda^0\,,\, \lambda^1)$ are the worldsheet coordinates parameterizing the string trajectory \cite{JLS, JP}, $\mathcal{M}$ is the string tension (a positive constant with dimensions of energy per unit length), and $\gamma$ is the determinant of the induced metric on the string worldsheet:

\begin{equation}\label{induced_metric}
\gamma_{ab} = g_{\mu\nu}\frac{\partial x^\mu}{\partial\lambda^a}\frac{\partial x^\nu}{\partial\lambda^b}.
\end{equation}
The antisymmetric tensor $\Sigma^{\mu\nu}$ in Eq. (\ref{NambuGoto}) represents the bi-vector density of the string cloud \cite{PSL}, defined as:
\begin{equation}
\Sigma^{\mu\nu} = \epsilon^{ab}\frac{\partial x^\mu}{\partial\lambda^a}\frac{\partial x^\nu}{\partial\lambda^b},
\end{equation}
where $\epsilon^{ab}$ is the two-dimensional Levi-Civita symbol. This formalism provides a relativistic description of string-like matter distributions in curved spacetime \cite{AV}.
The corresponding energy-momentum tensor takes the form \cite{PSL,DVS,EH}:
\begin{equation}\label{EMT}
T_{\mu\nu}^\text{cs} = \frac{\rho \Sigma_{\mu\sigma}\Sigma_{\nu}^{\;\sigma}}{\sqrt{-\gamma}},
\end{equation}
where the string bivector $\Sigma^{\mu\nu} = \epsilon^{ab}\partial_a x^\mu \partial_b x^\nu$ satisfies the conservation laws:
\begin{equation}\label{bivector_eqs}
\Sigma^{\mu\beta}\nabla_\mu\left[\frac{\Sigma_\beta^{\;\nu}}{(-\gamma)^{1/2}}\right] = 0, \quad \nabla_\mu(\rho\Sigma^{\mu\sigma})\Sigma_\sigma^{\;\nu} = 0.
\end{equation}The variation of the action \eqref{action} with metric $g_{\mu\nu}$ yields Einstein's equations with a nonzero cosmological constant are
\begin{equation}\label{EE}
R_{\mu\nu}-\frac{1}{2}g_{\mu\nu}R-3\,\ell^{-2}_p g_{\mu\nu}=T_{\mu\nu}^\mathrm{cs}.
\end{equation}The original Letelier solution \cite{PSL} considered only the electric-like component ($\Sigma_{01}$) while maintaining $\gamma < 0$. The  incorporation of both electric ($\Sigma_{01}$) and magnetic-like ($\Sigma_{23}$) components, yielding the spherically symmetric metric:

\begin{equation}\label{NCoS_metric}
ds^2 = -F(r)\,dt^2 +\frac{1}{F(r)}\,dr^2 +r^2\,(d\theta^2 + \sin^2\theta\, d\phi^2),
\end{equation}
with
\begin{equation}\label{metric_function}
F(r) =1 - \frac{2\,M}{r} + \frac{\lvert \alpha\rvert\, b^2}{r^2}\,{}_2F_1\left(-\frac{1}{2},-\frac{1}{4},\frac{3}{4},-\frac{r^4}{b^4}\right)+\frac{r^2}{\ell^2_p}.
\end{equation}
Here, $M$ means the BH mass, while $(\lvert \alpha\rvert,\,b)$ represent CoS parameters.

On the other hand, the study of the quintessence-like fluid as a matter content within general relativity was carried out by Kiselev~\cite{VVK}. He obtained a generalization of the Schwarzschild solution describing a BH surrounded by a quintessence-like fluid, with the corresponding energy-momentum tensor given by
\begin{equation}
    T^{t}_{t}=T^{r}_{r}=\rho_q,\quad\quad T^{\theta}_{\theta}=T^{\phi}_{\phi}=-\frac{1}{2}\,\rho_q\,(3\,w_q+1),\label{vvk}
\end{equation}
where $\rho_q$ denotes the energy density of the quintessence-like fluid, and the pressure is related to the density via the equation of state $p_q=w_q\,\rho_q$, with $w_q$ being the quintessence state parameter. 

Thereby, incorporating QF into the above BH solution, one can find a static and spherically symmetric AdS space-time with a CoS and surrounded by a QF describe by the following line element
\begin{equation}
ds^2 = -f(r)\,dt^2 +\frac{1}{f(r)}\,dr^2 + r^2\,(d\theta^2 + \sin^2\theta d\phi^2),\label{final}
\end{equation}
with the metric function is given by
\begin{equation}
f(r) =1 - \frac{2\,M}{r} + \frac{\lvert \alpha\rvert\, b^2}{r^2}\,{}_2F_1\left(-\frac{1}{2},-\frac{1}{4},\frac{3}{4},-\frac{r^4}{b^4}\right)-\frac{\mathrm{N}}{r^{3\,w+1}}+\frac{r^2}{\ell^2_p}\quad\quad (-1 < w <-1/3),\label{function}
\end{equation}
In this study, we consider this BH space-time model (\ref{final}) and analyze geodesic motions, thermodynamic topology, and scalar perturbations and analzye the outcomes.

In the limit $b \to 0$ and $\lvert \alpha\rvert = \alpha$, the metric function $f(r)$ reduces as,
\begin{equation}
f(r) =1 -\alpha- \frac{2\,M}{r}-\frac{\mathrm{N}}{r^{3\,w+1}}+\frac{r^2}{\ell^2_p},\label{function-2}
\end{equation}
In that limit, the solution (\ref{final}) reduces to the classical Letelier-AdS space-time embedded with quintessence-like fluid, which was discussed in \cite{MMDC}. Moreover, in the limit $\alpha \to 0$, the solution (\ref{final}) reduces to the Kiselev AdS BH solution \cite{VVK}, which is the generalization of Schwrazschild AdS BH immersed in quintessence-like fluid. Several recent investigations have explored BH solutions involving a CoS and QF within various gravitational frameworks. Notably, Ahmed {\it et al.} have examined different aspects of such systems, including geodesic motion, thermodynamics, and perturbative dynamics \cite{FA1,FA2,FA3,FA4,FA5,FA6,FA7,FA8}. These works collectively contribute to understanding how CoS and QF affect the geometry, stability, and observable features of BHs, offering insights into their physical and thermodynamic behavior in both classical and extended gravity theories.

\begin{figure}[ht!]
    \centering
    \includegraphics[width=0.5\linewidth]{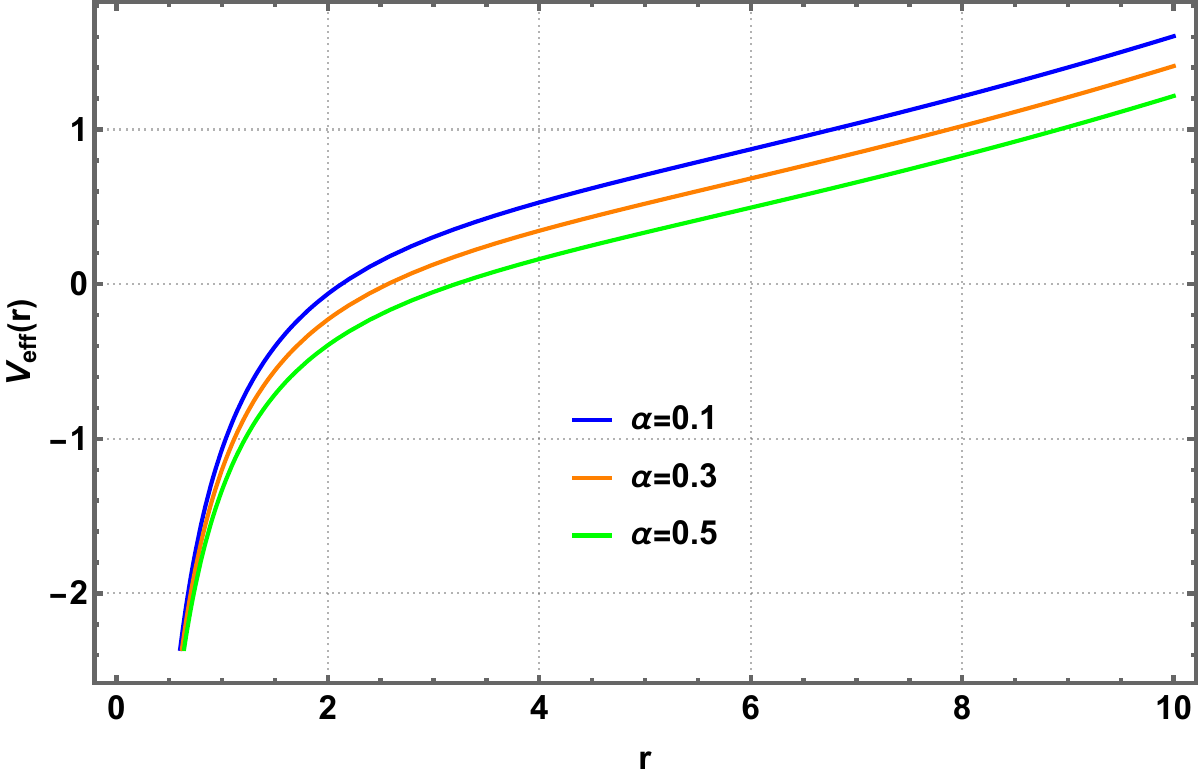}
    \caption{\footnotesize Behavior of the metric function $f(r)$ by varying CoS parameter $\alpha$. Here, we select $M=1, b=0.2, \ell_p=10, \mathrm{N}=0.01$.}
    \label{fig:function}
\end{figure}

Figure \ref{fig:function} illustrate the behavior of the metric function $f(r)$ given in Eq.~(\ref{function}) for different values of CoS parameter $\alpha$ with a particular state parameter $w=-2/3$. Here, we select the BH mass $M=1$, the CoS parameter $b=0.2$, the AdS radius $\ell_p=10$, and the normalization constant $\mathrm{N}=0.01$ of QF.

\section{Geodesics Structure}\label{sec:3}

In this section, we study geodesic motion of test particles (massless and massive) around the BH solution. We particularly focus into the photon trajectory, photon sphere radius, and BH shadow for massless particles, for example photons. Moreover, we aim to study dynamics of test particles, focusing into the innermost stable circular obits (ISCO). Recent studies of geodesics structure in various BH configurations can be found in \cite{FA1,FA2,FA3,FA4,FA5,FA6,FA7,FA8} and related references therein.

One of the common approach in investigating geodesic motion is the Lagrangian method which is defined by $\mathcal{L}=\frac{1}{2}\,g_{\mu\nu}\,\dot{x}^{\mu}\,\dot{x}^{\nu}$, where dot represents ordinary derivative w. r. to an affine parameter along the geodesic paths. Moreover, as the chosen space-time is spherically symmetric, we consider the geodesic motion in the equatorial plane defined by $\theta=\pi/2$ and $\dot{\theta}=0$. Expressing the chosen space-time (\ref{final}) in the form $ds^2=g_{\mu\nu}\,dx^{\mu}\,dx^{\nu}$, where $\mu, \nu=0,1,2,3$, the Lagrangian density function simplifies as:

\begin{equation}
    \mathcal{L}=\frac{1}{2}\,\left[-f(r)\,\dot{t}^2+\frac{1}{f(r)}\,\dot{r}^2+r^2\,\dot{\phi}^2\right].\label{bb1}
\end{equation}
As the space-time is static and spherically symmetric, there exist two Killing vector fields associated with the temporal coordinate $t$ denoted by $\xi_{(t)} \equiv \partial_{t}$ and the rotational symmetry associated with the azimuthal coordinate $\phi$ denoted by $\xi_{(\phi)} \equiv \partial_{\phi}$. The corresponding conserved quantities associated with these cyclic coordinates are the conserved energy defined by $-\mathrm{E}=g_{t\nu}\,\dot{x}^{\nu}$ and the conserved angular momentum $\mathrm{L}=g_{\phi \nu}\,\dot{x}^{\nu}$. In our case, these are explicitly given as,
\begin{equation}
    \mathrm{E}=f(r)\,\dot{t}.\label{bb2}
\end{equation}
And
\begin{equation}
    \mathrm{L}=r^2\,\dot{\phi}.\label{bb3}
\end{equation}

Eliminating $\dot{t}$ and $\dot{\phi}$ using Eqs.~(\ref{bb2}) and (\ref{bb3}) into the Eq. (\ref{bb1}), we find the equation of motion associated with the radial coordinate $r$ as,
\begin{equation}
    \dot{r}^2+V_\text{eff}(r)=\mathrm{E}^2\label{bb4}
\end{equation}
which is equivalent to the one-dimensional equation of motion of unit mass particles having energy $\mathrm{E}^2$ and the effective potential $V_\text{eff}(r)$, which governs the dynamics of massless and massive particles. This effective potential is given by
\begin{equation}
    V_\text{eff}(r)=\left(-\varepsilon+\frac{\mathrm{L}^2}{r^2}\right)\,f(r).\label{bb5}
\end{equation}
Here $\epsilon=0$ corresponds to the null geodesics and $-1$ for time-like.

From expression (\ref{bb5}), it becomes obvious that the effective potential for geodesic motion is influenced by the geometric and physical parameters. These include CoS parameters $(|\alpha|,\,b)$, the quintessence-like fluid parameters $(\mathrm{N}, w)$, the BH mass $M$, and the AdS radius $\ell_p$. Collectively these parameters alter the space-time curvature and shape the effective potential governing the dynamics of test particles in the gravitational field.

\begin{figure}[ht!]
    \centering
    \includegraphics[width=0.45\linewidth]{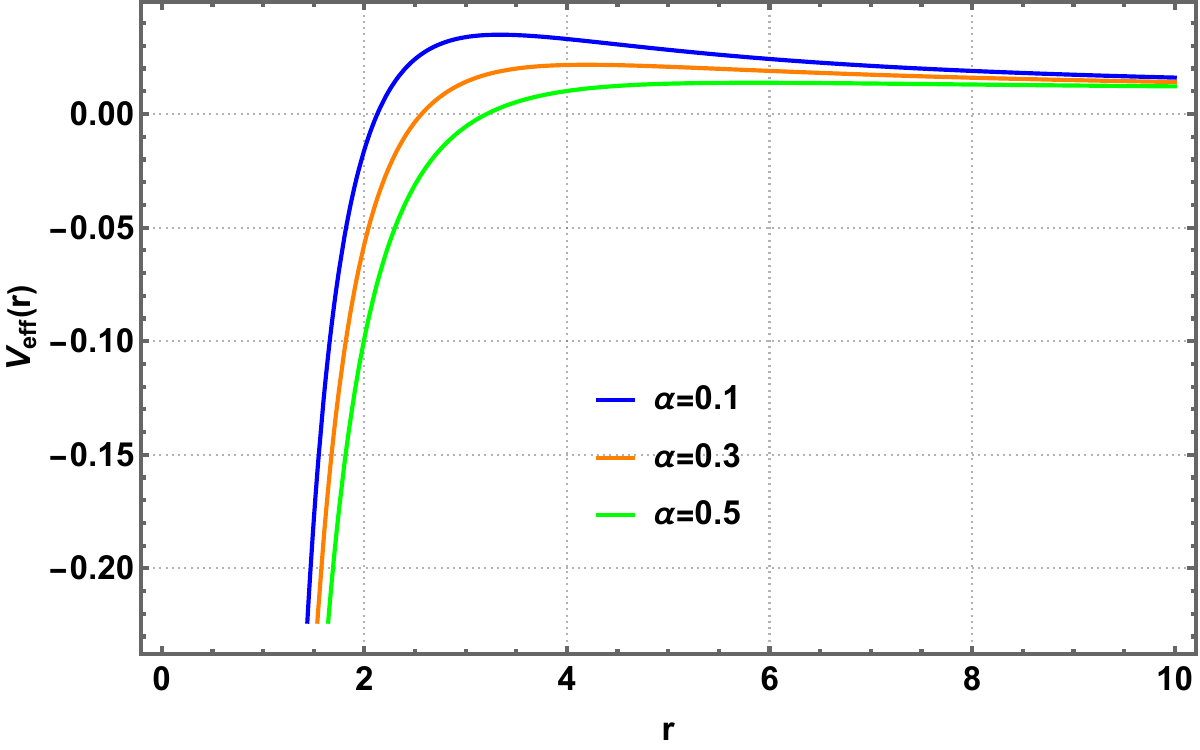}\quad\quad\quad
    \includegraphics[width=0.45\linewidth]{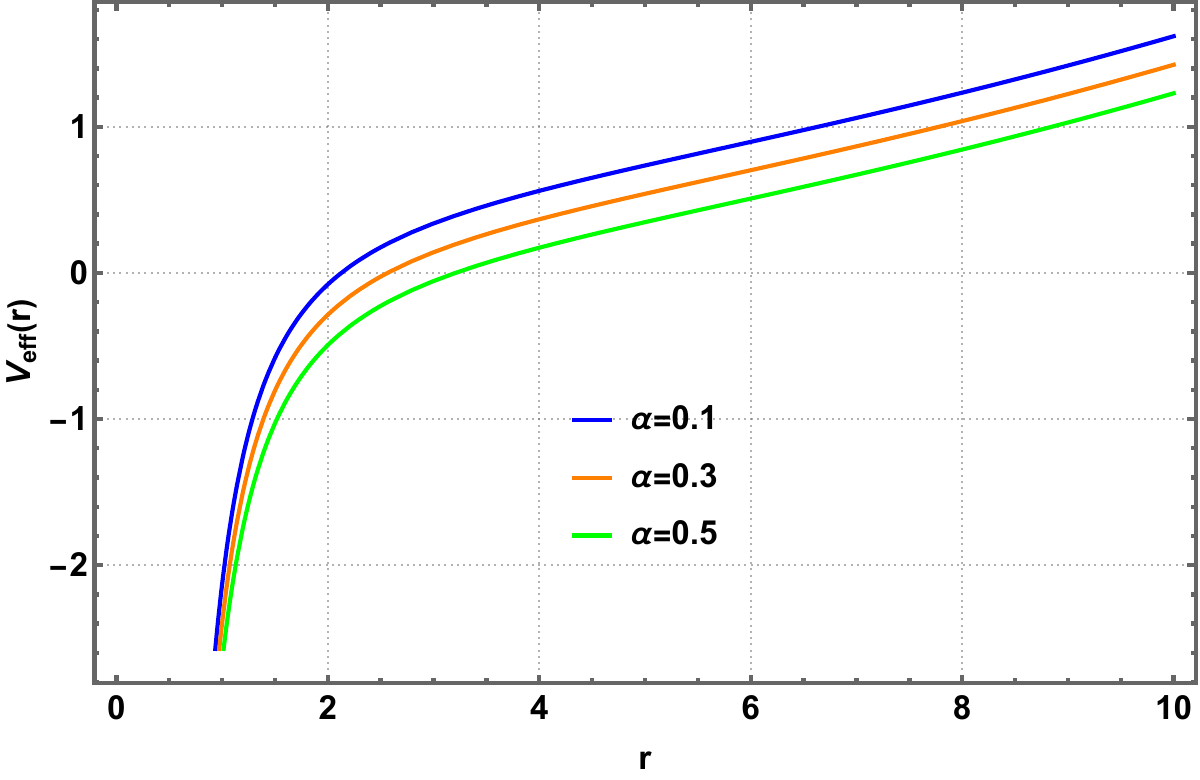}\\
    (a) $\varepsilon=0$ \hspace{6cm} (b) $\varepsilon=-1$
    \caption{\footnotesize Behavior of the effective potential for null and time-like geodesics by varying CoS parameter $\alpha$. Here, we set $M=1, b=0.2, \ell_p=10, w=-2/3, \mathrm{N}=0.01, \mathrm{L}=1$.}
    \label{fig:potential}
\end{figure}

\vspace{0.2cm}
\begin{center}
    \large{\bf A.\,Photon Dynamics}
\end{center}

In this part, we study the photon dynamics and analyze the photon trajectories, photon sphere radius, BH shadow. The photon dynamics is governed by the effective potential for null geodesic.

Thus, the effective potential from Eq. (\ref{bb5}) for null geodesic reduces as,
\begin{equation}
    V_\text{eff}(r)=\frac{\mathrm{L}^2}{r^2}\,f(r).\label{cc1}
\end{equation}

For circular null orbits, the conditions $\dot{r}=0$ and $\ddot{r}=0$ must be satisfied. Using Eq. (\ref{bb4}), these conditions simplifies to the following relations
\begin{equation}
    \mathrm{E}^2=V_\text{eff}(r)=\frac{\mathrm{L}^2}{r^2}\,f(r).\label{cc2}
\end{equation}
And
\begin{equation}
    V'_\text{eff}(r)=0,\label{cc3}
\end{equation}
where prime denotes partial derivative w. r. to $r$. 

The first relation (\ref{cc2}) gives us the critical impact parameter for photon particles and is given by
\begin{equation}
    \beta_c=\frac{r}{\sqrt{1 - \frac{2\,M}{r} + \frac{\lvert \alpha\rvert\, b^2}{r^2}\,{}_2F_1\left(-\frac{1}{2},-\frac{1}{4},\frac{3}{4},-\frac{r^4}{b^4}\right)-\frac{\mathrm{N}}{r^{3\,w+1}}+\frac{r^2}{\ell^2_p}}}.\label{cc4}
\end{equation}

For a specific state parameter, $w=-2/3$, the critical impact parameter simplifies as,
\begin{equation}
    \beta_c=\frac{r}{\sqrt{1 - \frac{2\,M}{r} + \frac{\lvert \alpha\rvert\, b^2}{r^2}\,{}_2F_1\left(-\frac{1}{2},-\frac{1}{4},\frac{3}{4},-\frac{r^4}{b^4}\right)-\mathrm{N}\,r+\frac{r^2}{\ell^2_p}}}.\label{cc5}
\end{equation}

Using the give metric function, we find $2\,f(r)-r\,f'(r)$ which is useful throughout the study as follows:
\begin{align}
   \frac{2\,f(r)-r\,f'(r)}{2}=1 - \frac{3\,M}{r} + \frac{2\,|\alpha|\, b^2}{r^2}\,{}_2F_1\left(-\frac{1}{2}, -\frac{1}{4}, \frac{3}{4}, -\frac{r^4}{b^4} \right) + \frac{|\alpha|\,r^2}{3\,b^2}\, {}_2F_1\left( \frac{1}{2}, \frac{3}{4}, \frac{7}{4}, -\frac{r^4}{b^4} \right) - \frac{(3\,w+3)\,\mathrm{N}}{2\,r^{3\,w+1}}.\label{special} 
\end{align}

The second relation (\ref{cc2}) using (\ref{cc1}) simplifies as,
\begin{equation}
    2\,f(r)-r\,f'(r)=0.\label{cc6}
\end{equation}
The above equation (\ref{cc6}) gives us the radius $r=r_s$ of the photon sphere. Substituting Eq.~(\ref{special}) into the above Eq.~(\ref{cc6}) results
\begin{align}
   1 - \frac{3\,M}{r} + \frac{2\,|\alpha|\, b^2}{r^2}\,{}_2F_1\left(-\frac{1}{2}, -\frac{1}{4}, \frac{3}{4}, -\frac{r^4}{b^4} \right) + \frac{|\alpha|\,r^2}{3\,b^2}\, {}_2F_1\left( \frac{1}{2}, \frac{3}{4}, \frac{7}{4}, -\frac{r^4}{b^4} \right) - \frac{(3\,w+3)\,\mathrm{N}}{2\,r^{3\,w+1}}=0.\label{cc7} 
\end{align}
From Eq.~(\ref{cc7}), it becomes clear that the photon sphere radius $r=r_s$ is influenced by the geometric and physical parameters. These include CoS parameters $(|\alpha|,\,b)$, the quintessence-like fluid parameters $(\mathrm{N}, w)$, the BH mass $M$, and the AdS radius $\ell_p$.

For a specific state parameter, $w=-2/3$, from Eq.~(\ref{cc7}), we find the following equation:
\begin{align}
   1 - \frac{3\,M}{r} + \frac{2\,|\alpha|\, b^2}{r^2}\,{}_2F_1\left(-\frac{1}{2}, -\frac{1}{4}, \frac{3}{4}, -\frac{r^4}{b^4} \right) + \frac{|\alpha|\,r^2}{3\,b^2}\, {}_2F_1\left( \frac{1}{2}, \frac{3}{4}, \frac{7}{4}, -\frac{r^4}{b^4} \right) -\frac{\mathrm{N}}{2}\,r=0.\label{cc8} 
\end{align}

From the above equation, we observes that for a particular state parameter $w=-2/3$, an analytical expression for the photon sphere radius $r=r_s$ is quite a challenging and difficult task. However, one can numerically determine this radius by choosing suitable values of other parameters involved in that equation (\ref{cc8}). In Table \ref{tab:1}, we present numerical values of the photon sphere radius $r_s$ by varying CoS parameter \(\alpha, b\) and the normalization constant $\mathrm{N}$ of QF for a particular state parameter $w=-2/3$.

\begin{table}[htbp]
\centering
\begin{minipage}{0.3\textwidth}
\centering
\begin{tabular}{|c|c|c|c|}
\hline
$|\alpha|$ & $b$ & $\mathrm{N}$ & $r_s$ \\
\hline
0.1 & 0.1 & 0.01 & 3.36812 \\
0.1 & 0.1 & 0.02 & 3.43629 \\
0.2 & 0.1 & 0.01 & 3.77555 \\
0.2 & 0.1 & 0.02 & 3.87406 \\
0.3 & 0.1 & 0.01 & 4.30944 \\
0.3 & 0.1 & 0.02 & 4.46109 \\
0.4 & 0.1 & 0.01 & 5.04244 \\
0.4 & 0.1 & 0.02 & 5.29845 \\
0.5 & 0.1 & 0.01 & 6.12043 \\
0.5 & 0.1 & 0.02 & 6.62316 \\
0.6 & 0.1 & 0.01 & 7.8986 \\
0.6 & 0.1 & 0.02 & 9.26455 \\
0.7 & 0.1 & 0.01 & 11.6811 \\
0.7 & 0.1 & 0.02 & 16.799 \\
0.8 & 0.1 & 0.01 & 23.6503 \\
0.8 & 0.1 & 0.02 & 16.7232 \\
0.9 & 0.1 & 0.01 & 23.5432 \\
0.9 & 0.1 & 0.02 & 16.647 \\
\hline
\end{tabular}\\
(i) $b = 0.1$
\end{minipage}
\hfill
\begin{minipage}{0.3\textwidth}
\centering
\begin{tabular}{|c|c|c|c|}
\hline
$|\alpha|$ & $b$ & $\mathrm{N}$ & $r_s$ \\
\hline
0.1 & 0.2 & 0.01 & 3.33878 \\
0.1 & 0.2 & 0.02 & 3.40573 \\
0.2 & 0.2 & 0.01 & 3.70889 \\
0.2 & 0.2 & 0.02 & 3.80377 \\
0.3 & 0.2 & 0.01 & 4.19346 \\
0.3 & 0.2 & 0.02 & 4.3365 \\
0.4 & 0.2 & 0.01 & 4.85776 \\
0.4 & 0.2 & 0.02 & 5.09351 \\
0.5 & 0.2 & 0.01 & 5.83176 \\
0.5 & 0.2 & 0.02 & 6.28058 \\
0.6 & 0.2 & 0.01 & 7.42701 \\
0.6 & 0.2 & 0.02 & 8.57635 \\
0.7 & 0.2 & 0.01 & 10.7343 \\
0.7 & 0.2 & 0.02 & 16.2608 \\
0.8 & 0.2 & 0.01 & 22.7743 \\
0.8 & 0.2 & 0.02 & 16.1038 \\
0.9 & 0.2 & 0.01 & 22.55 \\
0.9 & 0.2 & 0.02 & 15.9453 \\
\hline
\end{tabular}\\
(ii) $b = 0.2$
\end{minipage}
\hfill
\begin{minipage}{0.3\textwidth}
\centering
\begin{tabular}{|c|c|c|c|}
\hline
$|\alpha|$ & $b$ & $\mathrm{N}$ & $r_s$ \\
\hline
0.1 & 0.3 & 0.01 & 3.30945 \\
0.1 & 0.3 & 0.02 & 3.37518 \\
0.2 & 0.3 & 0.01 & 3.64227 \\
0.2 & 0.3 & 0.02 & 3.73361 \\
0.3 & 0.3 & 0.01 & 4.07767 \\
0.3 & 0.3 & 0.02 & 4.2124 \\
0.4 & 0.3 & 0.01 & 4.67367 \\
0.4 & 0.3 & 0.02 & 4.89022 \\
0.5 & 0.3 & 0.01 & 5.54495 \\
0.5 & 0.3 & 0.02 & 5.94415 \\
0.6 & 0.3 & 0.01 & 6.96214 \\
0.6 & 0.3 & 0.02 & 7.9273 \\
0.7 & 0.3 & 0.01 & 9.83197 \\
0.7 & 0.3 & 0.02 & 15.7043 \\
0.8 & 0.3 & 0.01 & 21.8633 \\
0.8 & 0.3 & 0.02 & 15.4596 \\
0.9 & 0.3 & 0.01 & 21.5118 \\
0.9 & 0.3 & 0.02 & 15.2111 \\
\hline
\end{tabular}\\
(iii) $b = 0.3$
\end{minipage}
\caption{\footnotesize Numerical results for the photon sphere radius $r=r_s$ using Eq.~(\ref{cc8}) for different values of CoS parameters $(|\alpha|,\,b)$ and the normalization constant $\mathrm{N}$ of QF. Here, we set $M=1$.}
\label{tab:1}
\end{table}

Next, we focus into the BH shadows cast by the BH solution, and examine how the geometric and physical parameters influence the size of the shadow. The shadow size $ R_s $ of a spherically symmetric BH (as seen by a distant observer) is determined by the photon sphere radius $ r_s$ and is typically expressed using the critical impact parameter $ \beta_c $. The radius of the shadow can be determined using the following formula:
\begin{equation}
R_s=\beta_c=\frac{r_s}{\sqrt{f(r_s)}}.\label{cc9}  
\end{equation}

For the space-time Eq.~(\ref{final}), the shadow size can be expressed as,
\begin{equation}
R_s=\frac{r_s}{\sqrt{1 - \frac{2\,M}{r_s} + \frac{\lvert \alpha\rvert\, b^2}{r^2_s}\,{}_2F_1\left(-\frac{1}{2},-\frac{1}{4},\frac{3}{4},-\frac{r^4_s}{b^4}\right)-\frac{\mathrm{N}}{r^{3\,w+1}_s}+\frac{r^2_s}{\ell^2_p}}}.\label{cc10}
\end{equation}

From Eq.~(\ref{cc10}), it becomes clear that the shadow size $R_s$ is influenced by several geometric and physical parameters. These include CoS parameters $(|\alpha|,\,b)$, the quintessence-like fluid parameters $(\mathrm{N}, w)$, the BH mass $M$, and the AdS radius $\ell_p$. In Table \ref{tab:2}, we present numerical values of both the photon sphere radius $r_s$ and shadow radius $R_s$. 

For a specific state parameter, $w=-2/3$, from Eq.~(\ref{cc10}), we find the shadow radius
\begin{equation}
R_s=\frac{r_s}{\sqrt{1 - \frac{2\,M}{r_s} + \frac{\lvert \alpha\rvert\, b^2}{r^2_s}\,{}_2F_1\left(-\frac{1}{2},-\frac{1}{4},\frac{3}{4},-\frac{r^4_s}{b^4}\right)-\mathrm{N}\,r_s+\frac{r^2_s}{\ell^2_p}}}.\label{cc11}
\end{equation}

\begin{table}[htbp]
\centering
\begin{minipage}[t]{0.32\textwidth}
\centering
\begin{tabular}{|c|c|c|c|c|}
\hline
$\alpha$ & $b$ & $\mathrm{N}$ & $r_s$ & $R_s$ \\
\hline
0.1 & 0.1 & 0.01 & 3.368 & 5.386 \\
0.2 & 0.1 & 0.01 & 3.776 & 6.092 \\
0.3 & 0.1 & 0.01 & 4.309 & 6.898 \\
0.4 & 0.1 & 0.01 & 5.042 & 7.775 \\
0.5 & 0.1 & 0.01 & 6.120 & 8.652 \\
0.6 & 0.1 & 0.01 & 7.899 & 9.410 \\
0.7 & 0.1 & 0.01 & 11.681 & 9.920 \\
0.8 & 0.1 & 0.01 & 23.650 & 10.105 \\
0.9 & 0.1 & 0.01 & 23.543 & 10.199 \\
\hline
\end{tabular}\\
(i) $b = 0.1$
\end{minipage}
\hfill
\begin{minipage}[t]{0.32\textwidth}
\centering
\begin{tabular}{|c|c|c|c|c|}
\hline
$\alpha$ & $b$ & $\mathrm{N}$ & $r_s$ & $R_s$ \\
\hline
0.1 & 0.2 & 0.01 & 3.339 & 5.352 \\
0.2 & 0.2 & 0.01 & 3.709 & 6.021 \\
0.3 & 0.2 & 0.01 & 4.193 & 6.792 \\
0.4 & 0.2 & 0.01 & 4.858 & 7.647 \\
0.5 & 0.2 & 0.01 & 5.832 & 8.526 \\
0.6 & 0.2 & 0.01 & 7.427 & 9.317 \\
0.7 & 0.2 & 0.01 & 10.734 & 9.879 \\
0.8 & 0.2 & 0.01 & 22.774 & 10.101 \\
0.9 & 0.2 & 0.01 & 22.550 & 10.203 \\
\hline
\end{tabular}\\
(ii) $b = 0.2$
\end{minipage}
\hfill
\begin{minipage}[t]{0.32\textwidth}
\centering
\begin{tabular}{|c|c|c|c|c|}
\hline
$\alpha$ & $b$ & $\mathrm{N}$ & $r_s$ & $R_s$ \\
\hline
0.1 & 0.3 & 0.01 & 3.309 & 5.317 \\
0.2 & 0.3 & 0.01 & 3.642 & 5.947 \\
0.3 & 0.3 & 0.01 & 4.078 & 6.682 \\
0.4 & 0.3 & 0.01 & 4.674 & 7.510 \\
0.5 & 0.3 & 0.01 & 5.545 & 8.386 \\
0.6 & 0.3 & 0.01 & 6.962 & 9.208 \\
0.7 & 0.3 & 0.01 & 9.832 & 9.826 \\
0.8 & 0.3 & 0.01 & 21.863 & 10.097 \\
0.9 & 0.3 & 0.01 & 21.512 & 10.208 \\
\hline
\end{tabular}\\
(iii) $b = 0.3$
\end{minipage}
\caption{\footnotesize Numerical results for the photon sphere $r_s$ and shadow radius $R_s$ for various values of $\alpha$, with different $b$ values and fixed $\mathrm{N} = 0.01, w=-2/3.$}
\label{tab:2}
\end{table}

Beyond the classical approach in determining the photon sphere radius, a topological viewpoint offers additional insight by assigning a topological charge to each photon orbit, encapsulating its stability traits. This approach involves defining the scalar function \cite{45'm,45mmm,45m,47m,48m,49m,50m}
\begin{equation}\label{p6}
H(r, \theta) = \sqrt{\frac{-g_{tt}}{g_{\phi\phi}}} = \frac{\sqrt{f(r)}}{r\,\sin \theta},
\end{equation}
where $g_{tt}$ and $g_{\phi\phi}$ are metric components, and $h(r)$ is another radial function characterizing the metric structure. The radii of photon spheres correspond to critical points of $H$ with respect to $r$, identified by the condition
\begin{equation}\label{p7}
\frac{dH}{dr} = 0.
\end{equation}
To further explore the topological structure, one constructs a vector field $\boldsymbol{\varphi}$ on the two-dimensional manifold spanned by $(r_h, \theta)$ coordinates, with components defined as
\begin{equation}\label{p8}
\varphi^{r_h} = \frac{1}{\sqrt{f(r)}}\, \frac{dH}{dr}, \quad \varphi^{\theta} = \frac{\partial_\theta H}{r}.
\end{equation}
This vector field can be expressed in complex or polar form as
\begin{equation}\label{p9}
\varphi = |\varphi| e^{i \Theta} = \varphi^{r_h} + i \varphi^{\theta},
\end{equation}
with magnitude
\begin{equation}\label{p10}
|\varphi| = \sqrt{(\varphi^{r_h})^2 + (\varphi^{\theta})^2}.
\end{equation}
Normalizing $\boldsymbol{\varphi}$ yields a unit vector field
\begin{equation}\label{p11}
n^{a} = \frac{\varphi^{a}}{|\varphi|},
\end{equation}
where the index $a = 1,2$ corresponds to directions along $r_h$ and $\theta$, respectively. For a deeper topological characterization of photon spheres, scalar fields $\phi^{r_h}$ and $\phi^{\Theta}$ are introduced. These fields capture aspects of stability and dynamics from a topological perspective, providing insight into the interplay between geometry, motion, and topological classification of photon spheres around BHs:
\begin{align}
\phi^{r_h} &= \frac{1}{2} \csc (\theta ) r^{-3\, \omega -4} \left(r \left(3\, n\, (\omega +1)-2\, (r-3\, M)\, r^{3\, \omega }\right)-b^2\, | \alpha |\,  r^{3\, \omega }\, \left(3 \, _2F_1\left(-\frac{1}{2},-\frac{1}{4};\frac{3}{4};-\frac{r^4}{b^4}\right)+\sqrt{\frac{r^4}{b^4}+1}\right)\right),\label{p12}\\
\phi^{\theta} &=-\frac{\cot (\theta )\, \csc (\theta )\, \sqrt{\frac{b^2\, | \alpha |}{r^2}\,{}_2F_1\left(-\frac{1}{2},-\frac{1}{4};\frac{3}{4};-\frac{r^4}{b^4}\right)-\frac{2\,M}{r}-n\, r^{-3\, \omega -1}+\frac{3\, r^2}{8\, \pi\,  P}+1}}{r^2}.\label{p13}
\end{align}

\begin{figure}[ht!]
 \begin{center}
 \subfigure[]{
 \includegraphics[height=3.5cm,width=3.8cm]{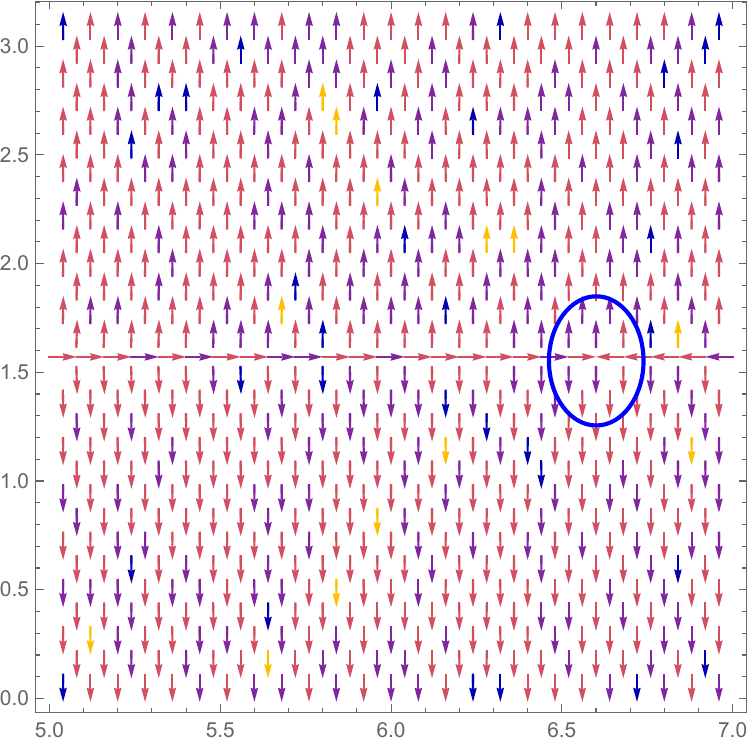}
 \label{200a}}\quad\quad
 \subfigure[]{
 \includegraphics[height=3.5cm,width=3.8cm]{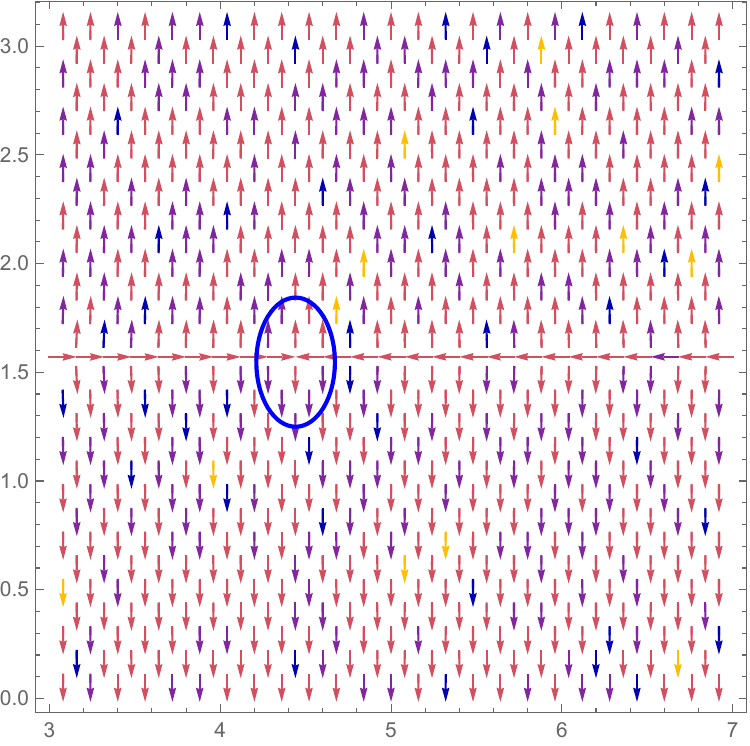}
 \label{200b}}\quad\quad
 \subfigure[]{
 \includegraphics[height=3.5cm,width=3.8cm]{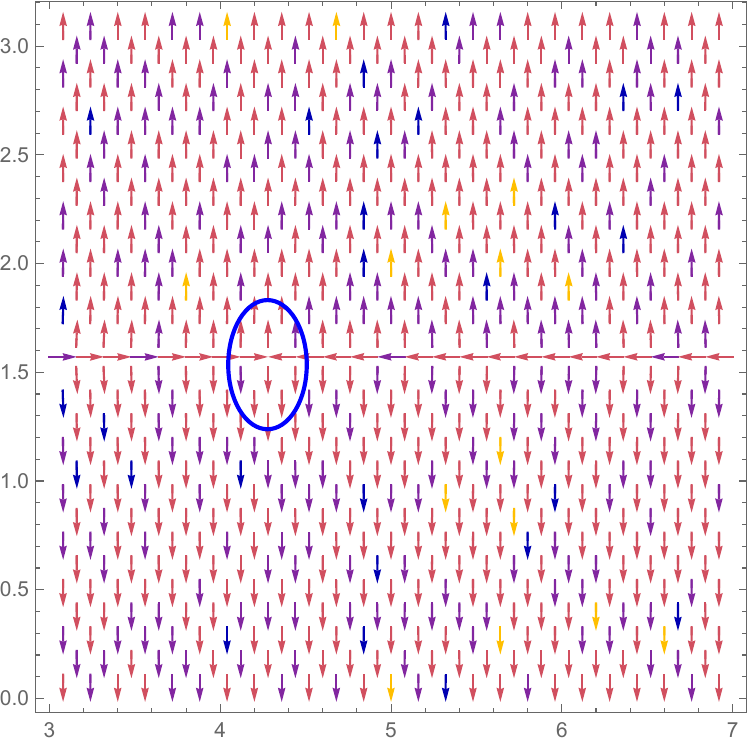}
 \label{200c}}\\
 \subfigure[]{
 \includegraphics[height=3.5cm,width=3.8cm]{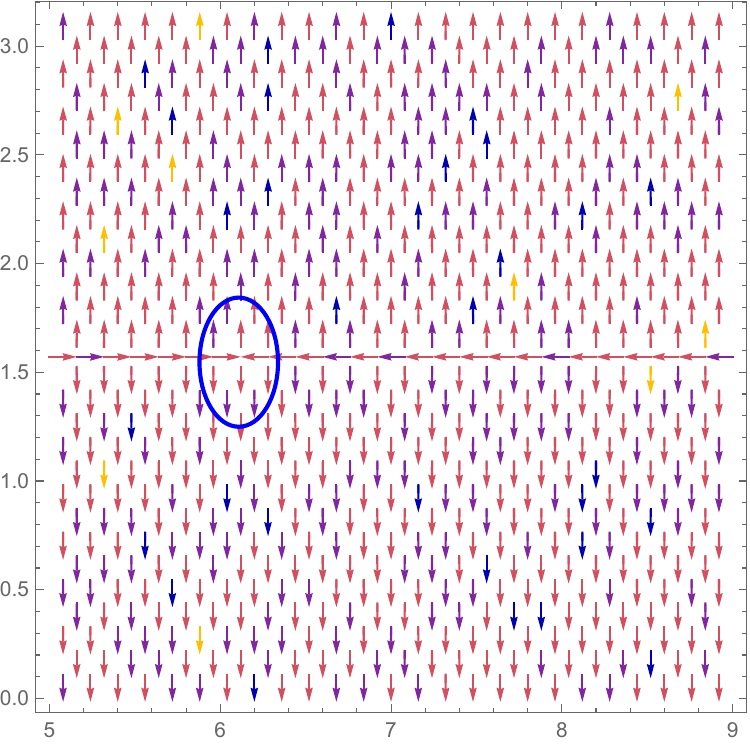}
 \label{200d}}\quad\quad
 \subfigure[]{
 \includegraphics[height=3.5cm,width=3.8cm]{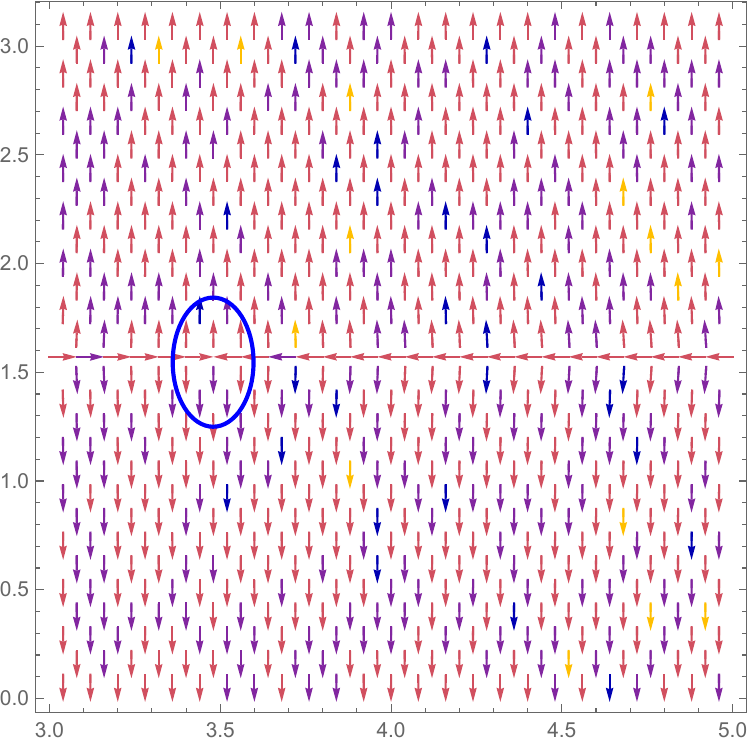}
 \label{200e}}\quad\quad
 \subfigure[]{
 \includegraphics[height=3.5cm,width=3.8cm]{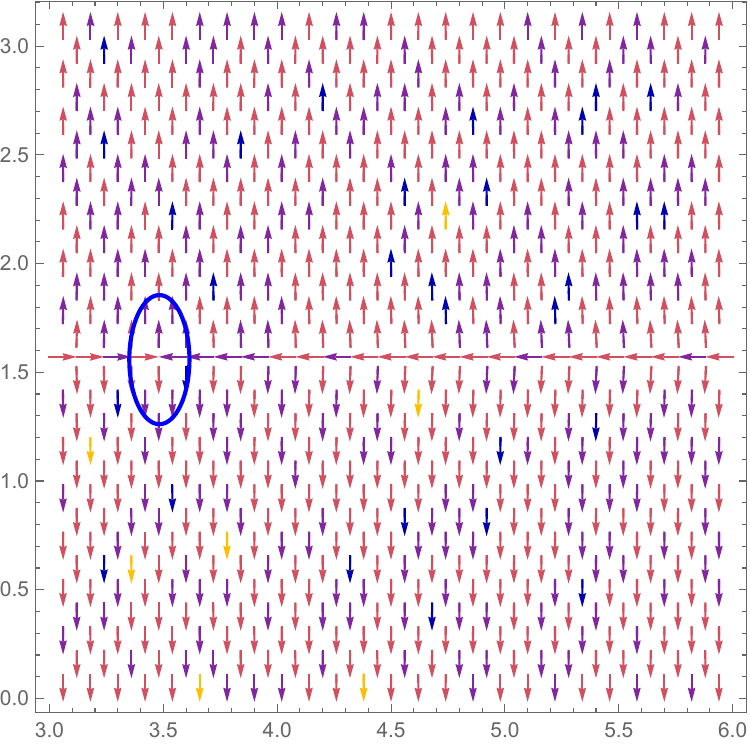}
 \label{200f}}
  \caption{\small{The following set of figures presents the photon sphere (PSs) structures obtained for different combinations of key parameters:
Figure (\ref{200a}) illustrates the photon sphere for $\alpha = 0.5$ and $b = 0.1$. Figure (\ref{200b}) shows the configuration when $\alpha = 0.3$ and $b = 0.15$. Figure (\ref{200c}) depicts the photon sphere corresponding to $\alpha = 0.3$ and $b = 0.25$. Figure (\ref{200d}) demonstrates the case with $\alpha = 0.5$ and $b = 0.25$. Figure (\ref{200e}) presents the structure for $\alpha = 0.1$ combined with $b = 0.05$. Figure (\ref{200f}) reveals the photon sphere arrangement at $\alpha = 0.1$ and $b = 0.1$. All these configurations are evaluated under the condition of a fixed charge parameter set as $\omega = -\frac{2}{3}$, along with a constant value of $N = 0.02$ and mass $M = 1$. Together, these images offer an in-depth depiction of how the photon sphere morphology responds to changes in the symmetry-breaking energy scale $\alpha$ and the coupling constant $b$, highlighting the sensitivity of photon trajectories and gravitational lensing effects to these parameters.}}
\label{T100}
\end{center}
 \end{figure}
 
Within the scope of thermodynamic topology, we undertake a comprehensive investigation of photon spheres (PSs) and their pivotal function in assessing the stability properties of BHs. A core topological concept states that each zero of the vector field $\phi$, when encircled by a closed contour, contributes a quantized topological charge defined by its winding number around that loop. In this context, every photon sphere is attributed a charge value of either $+1$ or $-1$, dictated by the orientation and the direction in which the vector field $\phi$ winds around the zero point. The cumulative topological charge within any given domain, which may encompass multiple zeros of $\phi$, takes on discrete values restricted to $-1$, $0$, or $+1$. For typical BH solutions satisfying $M > Q$, the overall topology of the photon spheres is generally characterized by a net charge of $-1$, reflecting the underlying geometric and physical constraints of the spacetime. To explore how these topological features evolve with changes in physical parameters, we systematically analyze the photon sphere configurations over a range of parameters: the symmetry-breaking scale parameter $\alpha = 0.1, 0.3, 0.5$; the coupling constant $b = 0.05, 0.1, 0.15, 0.25$; alongside fixed values of mass $M = 1$, normalization factor $N = 0.02$, and charge parameter $\omega = -\frac{2}{3}$. The outcomes of this parameter sweep are presented in Fig.~~\ref{T100}, which clearly demonstrates that the total topological charge associated with photon spheres remains invariant at $-1$ across the entire parameter space considered.

A thorough analysis of these photon sphere structures not only enriches our understanding of the BH’s effective geometry but also sheds light on their thermodynamic stability properties. By examining photon orbits and their corresponding effective potentials, we observe how the interplay between $\alpha$ and $b$ modulates the topological characteristics inherent in the photon sphere configurations. Crucially, our study highlights the distinct signatures of unstable photon spheres, which serve as markers for phase transitions and instabilities within the BH system.

Moreover, the relationship between the topological charges of BHs and the associated photon sphere topology provides a profound link that extends beyond theoretical considerations. This connection has practical implications for astrophysical phenomena such as gravitational lensing and the formation of BH shadows, both of which depend sensitively on the photon sphere structure. The integration of these topological methods with observational physics offers promising avenues for testing the fundamental nature of gravity in strong-field regimes.

Now, we determine effective radial force experienced by the photon particles in the gravitational field. This force can be determined using the effective potential for null geodesics defined by $\mathcal{F}=-\frac{1}{2}\,\frac{dV_\text{eff}(r)}{dr}$. Using the effective potential given in Eq.~(\ref{cc1}), we find
\begin{equation}
    \mathcal{F}_\text{rad}=\frac{\mathrm{L}^2}{r^3}\,\left[1 - \frac{3\,M}{r} + \frac{2\,|\alpha|\, b^2}{r^2}\,{}_2F_1\left(-\frac{1}{2}, -\frac{1}{4}, \frac{3}{4}, -\frac{r^4}{b^4} \right) + \frac{|\alpha|\,r^2}{3\,b^2}\, {}_2F_1\left( \frac{1}{2}, \frac{3}{4}, \frac{7}{4}, -\frac{r^4}{b^4} \right) - \frac{(3\,w+3)\,\mathrm{N}}{2\,r^{3\,w+1}}\right].\label{cc12}
\end{equation}

From Eq.~(\ref{cc12}), it becomes clear that the effective radial force experienced by the photon particles is influenced by several geometric and physical parameters. These include CoS parameters $(|\alpha|,\,b)$, the quintessence-like fluid parameters $(\mathrm{N}, w)$, the BH mass $M$, and the AdS radius $\ell_p$.

For a specific state parameter, $w=-2/3$, this radial force from Eq.~(\ref{cc12}) reduces as,
\begin{equation}
    \mathcal{F}_\text{rad}=\frac{\mathrm{L}^2}{r^2}\,\left[1 - \frac{3\,M}{r} + \frac{2\,|\alpha|\, b^2}{r^2}\,{}_2F_1\left(-\frac{1}{2}, -\frac{1}{4}, \frac{3}{4}, -\frac{r^4}{b^4} \right) + \frac{|\alpha|\,r^2}{3\,b^2}\, {}_2F_1\left( \frac{1}{2}, \frac{3}{4}, \frac{7}{4}, -\frac{r^4}{b^4} \right) - \frac{\mathrm{N}}{2}\,r\right].\label{cc13}
\end{equation}

\begin{figure}[ht!]
    \centering
    \includegraphics[width=0.5\linewidth]{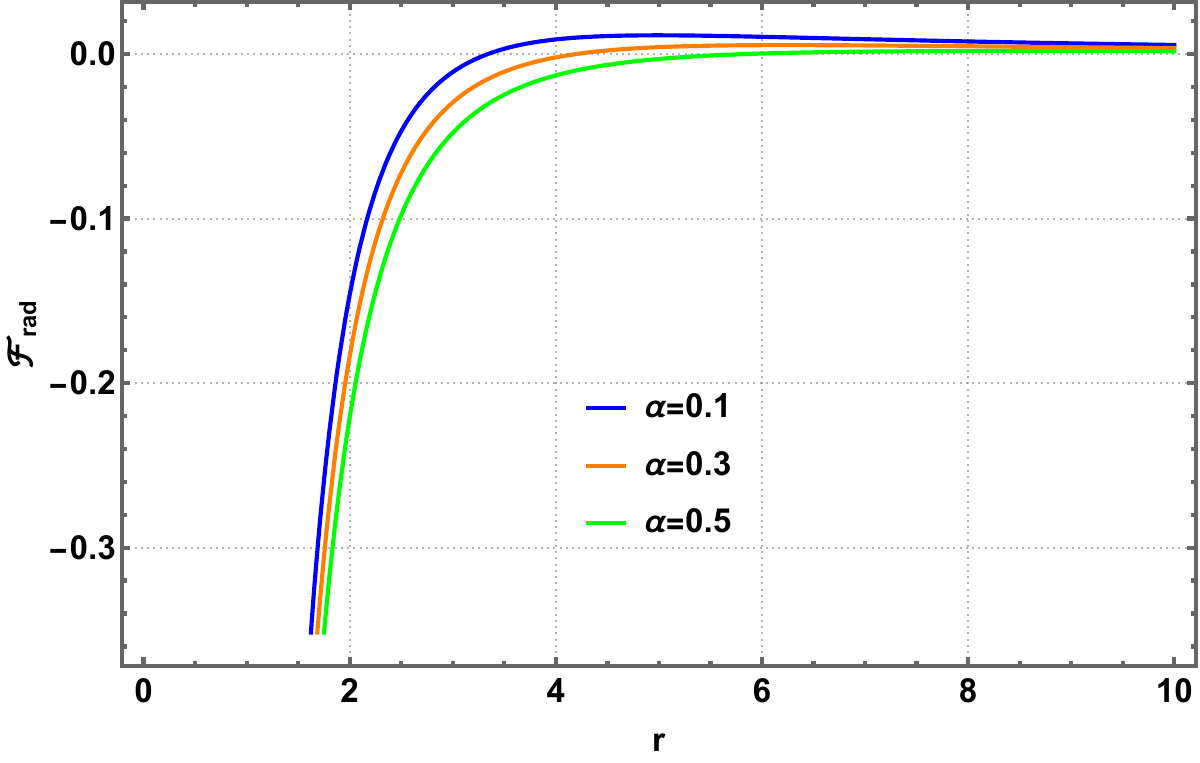}
    \caption{\footnotesize Behavior of the effective radial force by varying CoS parameter $\alpha$. Here, we set $M=1, b=0.2, \ell_p=10, w=-2/3, \mathrm{N}=0.01, \mathrm{L}=1$.}
    \label{fig:force}
\end{figure}

In Figure \ref{fig:force}, we illustrate the behavior of the effective radial force as a function of $r$ for different values of $\alpha=0.1,0.2,0.3$. It clear from the Figure that as the value of $\alpha$ increases, the radial force experienced by the photon particles reduces.

\vspace{0.2cm}
\begin{center}
    \large{\bf B.\,Test Particle Dynamics}
\end{center}

In this part, we study the test particle dynamics, specifically focus into the ISCO radius and show how several geometric and physical parameters influences this ISCO radius.

For massive test particles, the effective potential from Eq. (\ref{bb5}) reduces as
\begin{equation}
    V_\text{eff}(r)=\left(1+\frac{\mathrm{L}^2}{r^2}\right)\,f(r).\label{dd1}
\end{equation}

For circular orbits, the following conditions must be satisfied
\begin{align}
    &\mathrm{E}^2=V_\text{eff}(r)=\left(1+\frac{\mathrm{L}^2}{r^2}\right)\,f(r),\label{dd2}\\
    &\partial_{r}\,V_\text{eff}(r)=0,\label{dd3}\\
    &\partial_{rr}\,V_\text{eff}(r) \geq 0.\label{dd4}
\end{align}

Substituting the effective potential in condition (\ref{dd3}), we find the specific angular momentum of test particles as,
\begin{equation}
    \mathrm{L}_\text{specific}=\pm\,r\,\sqrt{\frac{\frac{M}{r} 
- \frac{|\alpha|\,b^{2}}{r^{2}} \, {}_2F_1\left(-\frac{1}{2}, -\frac{1}{4}, \frac{3}{4}, -\frac{r^{4}}{b^{4}}\right)- \frac{|\alpha|\, r^{2}}{3\,b^{2}} \, {}_2F_1\left(\frac{1}{2}, \frac{3}{4}, \frac{7}{4}, -\frac{r^{4}}{b^{4}}\right)+ \frac{\mathrm{N}\,(3\, w + 1)}{2\,r^{3\,w + 1}}+ \frac{r^{2}}{\ell_p^{2}}}{1 - \frac{3\,M}{r} + \frac{2\,|\alpha|\, b^2}{r^2}\,{}_2F_1\left(-\frac{1}{2}, -\frac{1}{4}, \frac{3}{4}, -\frac{r^4}{b^4} \right) + \frac{|\alpha|\,r^2}{3\,b^2}\, {}_2F_1\left( \frac{1}{2}, \frac{3}{4}, \frac{7}{4}, -\frac{r^4}{b^4} \right) - \frac{(3\,w+3)\,\mathrm{N}}{2\,r^{3\,w+1}}}}.\label{dd5}
\end{equation}
Thereby, using (\ref{dd5}) into the condition (\ref{dd2}), we find the specific energy of test particles as,
\begin{equation}
    \mathrm{E}_\text{specific}=\sqrt{\frac{1 - \frac{2\,M}{r} + \frac{\lvert \alpha\rvert\, b^2}{r^2}\,{}_2F_1\left(-\frac{1}{2},-\frac{1}{4},\frac{3}{4},-\frac{r^4}{b^4}\right)-\frac{\mathrm{N}}{r^{3\,w+1}}+\frac{r^2}{\ell^2_p}}{1 - \frac{3\,M}{r} + \frac{2\,|\alpha|\, b^2}{r^2}\,{}_2F_1\left(-\frac{1}{2}, -\frac{1}{4}, \frac{3}{4}, -\frac{r^4}{b^4} \right) + \frac{|\alpha|\,r^2}{3\,b^2}\, {}_2F_1\left( \frac{1}{2}, \frac{3}{4}, \frac{7}{4}, -\frac{r^4}{b^4} \right) - \frac{(3\,w+3)\,\mathrm{N}}{2\,r^{3\,w+1}}}}.\label{dd6}
\end{equation}

From Eqs.~(\ref{dd5}) and (\ref{dd6}), it becomes clear that the specific angular momentum and energy of test particles are influenced by geometric and physical parameters. These include CoS parameters $(|\alpha|,\,b)$, the quintessence-like fluid parameters $(\mathrm{N}, w)$, the BH mass $M$, and the AdS radius $\ell_p$.

For a specific state parameter, $w=-2/3$, these physical quantities reduces as
\begin{align}
    \mathrm{L}_\text{specific}&=\pm\,r\,\sqrt{\frac{\frac{M}{r} 
- \frac{|\alpha|\,b^{2}}{r^{2}} \, {}_2F_1\left(-\frac{1}{2}, -\frac{1}{4}, \frac{3}{4}, -\frac{r^{4}}{b^{4}}\right)- \frac{|\alpha|\, r^{2}}{3\,b^{2}} \, {}_2F_1\left(\frac{1}{2}, \frac{3}{4}, \frac{7}{4}, -\frac{r^{4}}{b^{4}}\right)-\frac{\mathrm{N}}{2}\,r+ \frac{r^{2}}{\ell_p^{2}}}{1 - \frac{3\,M}{r} + \frac{2\,|\alpha|\, b^2}{r^2}\,{}_2F_1\left(-\frac{1}{2}, -\frac{1}{4}, \frac{3}{4}, -\frac{r^4}{b^4} \right) + \frac{|\alpha|\,r^2}{3\,b^2}\, {}_2F_1\left( \frac{1}{2}, \frac{3}{4}, \frac{7}{4}, -\frac{r^4}{b^4} \right) - \frac{\mathrm{N}}{2}\,r}},\nonumber\\
\mathrm{E}_\text{specific}&=\sqrt{\frac{1 - \frac{2\,M}{r} + \frac{\lvert \alpha\rvert\, b^2}{r^2}\,{}_2F_1\left(-\frac{1}{2},-\frac{1}{4},\frac{3}{4},-\frac{r^4}{b^4}\right)-\frac{\mathrm{N}}{2}\,r+\frac{r^2}{\ell^2_p}}{1 - \frac{3\,M}{r} + \frac{2\,|\alpha|\, b^2}{r^2}\,{}_2F_1\left(-\frac{1}{2}, -\frac{1}{4}, \frac{3}{4}, -\frac{r^4}{b^4} \right) + \frac{|\alpha|\,r^2}{3\,b^2}\, {}_2F_1\left( \frac{1}{2}, \frac{3}{4}, \frac{7}{4}, -\frac{r^4}{b^4} \right) - \frac{\mathrm{N}}{2}\,r}}.\label{dd7}
\end{align}

Substituting the effective potential given in Eq.~(\ref{dd1}) into the  condition (\ref{dd4}) and using (\ref{dd3}), we find the following equation satisfying the ISCO radius $r=r_\text{ISCO}$ as (setting $w=-2/3$)
\begin{align}
& 3\,\left(1 - \frac{2\,M}{r} + \frac{|\alpha|\,b^{2}}{r^{2}}\,{}_2F_1- \mathrm{N}\,r+ \frac{r^{2}}{\ell_p^{2}} \right)\,\left[ \frac{2\,M}{r^{2}} + |\alpha|\, b^{2} \left(-2\,\frac{{}_2F_1}{r^{3}}+ \frac{{}_2F'_1}{r^{2}} \right) -\mathrm{N}+ \frac{2\,r}{\ell_p^{2}} \right] \nonumber\\
& \quad - 2\, r\, \left[ \frac{2\,M}{r^{2}} + |\alpha|\, b^{2} \left(-2\,\frac{\,{}_2F_1}{r^{3}}+ \frac{{}_2 F'_1}{r^{2}} \right) -\mathrm{N}+ \frac{2\,r}{\ell_p^{2}} \right]^2\nonumber\\ 
& \quad + r\, \left(1 - \frac{2\,M}{r} + \frac{|\alpha|\, b^{2}}{r^{2}}\,{}_2F_1- \mathrm{N}\,r+ \frac{r^{2}}{\ell_p^{2}} \right)\,\left[ - \frac{4\,M}{r^{3}} + |\alpha|\, b^{2} \left( \frac{{}_2F''_1}{r^{2}} -6\,\frac{{}_2F'_1}{r^{3}} +6\, \frac{{}_2 F_1}{r^{4}} \right)+ \frac{2}{\ell_p^{2}} \right] = 0.\label{dd8}
\end{align}

where
\[
\begin{cases}
{}_2F_1={}_2F_1\left(-\frac{1}{2}, -\frac{1}{4}, \frac{3}{4}, -\frac{r^4}{b^4} \right),\\[10pt]
{}_2F'_1= - \frac{3\,b^4}{2\,r^3} \, {}_2F_1\left(\frac{1}{2}, \frac{3}{4}; \frac{7}{4}; -\frac{r^4}{b^4}\right), \\[10pt]
{}_2F''_1= - \frac{b^4}{2\, r^2} \, {}_2F_1\left(\frac{1}{2}, \frac{3}{4}; \frac{7}{4}; -\frac{r^4}{b^4}\right) 
+ \frac{7\, b^8}{4\, r^6} \, {}_2F_1\left(\frac{3}{2}, \frac{7}{4}; \frac{11}{4}; -\frac{r^4}{b^4}\right).
\end{cases}
\]

From Eq.~(\ref{dd8}), it becomes clear that the ISCO radius depends on several geometric and physical parameters. These include CoS parameters $(|\alpha|,\,b)$, the quintessence-like fluid parameters $(\mathrm{N}, w)$, the BH mass $M$, and the AdS radius $\ell_p$. An analytical expression for ISCO radius $r=r_\text{ISCO}$ is quite a challenging task. However, numerically one can determine this radius by setting suitable values of parameters involved in the equation. In Table \ref{tab:3}, we presented numerical values ISCO radius for $M=1, \ell_p=10$ and $\mathrm{N}=0.01$.

\begin{table}[ht!]
\centering
\begin{tabular}{|c|c|c|c|c|c|}
\hline
$b(\downarrow) \backslash \alpha (\rightarrow)$ & 0.1 & 0.2 & 0.3 & 0.4 & 0.5 \\
\hline
0.1 & 6.41846 & 11.72612 & 18.55002 & 10.38415 & \text{No positive root} \\
\hline
0.2 & 6.33909 & 11.43015 & 18.46810 & 10.22061 & \text{No positive root} \\
\hline
0.3 & 6.25952 & 11.13615 & 18.38519 & 10.05119 & \text{No positive root} \\
\hline
\end{tabular}
\caption{\footnotesize Numerical results for the ISCO radius $r_\text{ISCO}$ for different values of CoS parameter $(|\alpha|,\,b)$, while keeping $\mathrm{N}=0.01, M=1, \ell_p=10$ fixed.}
\label{tab:3}
\end{table}

From Table \ref{tab:3}, we observe that there is no positive real root for the ISCO radius for $\alpha\geq 0.5$ whatever values of $b$. 

\section{Thermodynamic topology} \label{sec:4}

The framework of thermodynamic topology provides a robust mathematical toolset for exploring the topological aspects inherent to BH (BH) thermodynamics. This approach is essential for studying phase transitions, determining stability conditions, and classifying BH states via their associated topological charges and critical points. By rigorously characterizing these elements, it becomes possible to systematically organize BH phases and anticipate their behavior under various physical scenarios. To establish a formal classification, we adopt the F-method which introduces a generalized free energy function defined by \cite{a19, a20}:
\begin{equation}\label{F1}
\mathcal{F} = M - \frac{S}{\tau},
\end{equation}
where $\tau$ denotes the Euclidean time period, and the temperature $T$ of the ensemble is given by its inverse $T = 1/\tau$. Importantly, the free energy is evaluated on-shell by enforcing $\tau = \tau_h = 1/T_h$, with $T_h$ representing the Hawking temperature. Central to the topological characterization is the vector field $\phi$ formulated as \cite{a19, a20}
\begin{equation}\label{F2}
\phi = \left( \frac{\partial \mathcal{F}}{\partial r_h},\ -\cot \Theta \csc \Theta \right).
\end{equation}
Here, the first component measures the sensitivity of the free energy with respect to variations in the horizon radius $r_h$, while the second component introduces a nontrivial angular dependence crucial for forming a complete topological vector field. This structure enables the detection of critical points that correspond to distinct thermodynamic phases of the BH. Notably, the angular component $\phi^\Theta$ diverges at the boundaries $\Theta=0$ and $\Theta=\pi$, causing the vector field to extend outward at these limits. The physically relevant parameter domain for our investigation is constrained by $0 \leq r_h < \infty$ and $0 \leq \Theta \leq \pi$, capturing all meaningful BH configurations. To quantitatively describe the topology, we define the topological current:
\begin{equation}\label{F32}
j^{\mu} = \frac{1}{2\pi} \varepsilon^{\mu\nu\rho} \varepsilon_{ab} \partial_{\nu} n^{a} \partial_{\rho} n^{b}, \quad \mu,\nu,\rho = 0,1,2,
\end{equation}
where the normalized vector field $n$ is given by
\begin{equation}\label{F4}
n^1 = \frac{\phi^{r_h}}{|\phi|}, \quad n^2 = \frac{\phi^\Theta}{|\phi|}.
\end{equation}
This normalization procedure ensures that the resulting topological quantities depend solely on the directional behavior of the vector field rather than its magnitude. The conservation law for the topological current is expressed as \cite{a19, a20}:
\begin{equation}\label{F5}
\partial_{\mu} j^{\mu} = 0,
\end{equation}
signifying that the associated topological charges are conserved quantities. For explicit calculation of topological invariants, the current $j^\mu$ can be written in terms of the delta function localized at zeros of $\phi$:
\begin{equation}\label{F6}
j^{\mu} = \delta^{2}(\phi) J^{\mu}\left(\frac{\phi}{x}\right),
\end{equation}
where the Jacobi tensor $J^\mu(\phi/x)$ satisfies
\begin{equation}\label{F7}
\varepsilon^{ab} J^{\mu} \left(\frac{\phi}{x}\right) = \varepsilon^{\mu\nu\rho} \partial_{\nu} \phi^a \partial_{\rho} \phi^b.
\end{equation}
For the time component $\mu = 0$, the Jacobian reduces to the determinant
\begin{equation}\label{F8}
J^0 \left(\frac{\phi}{x}\right) = \frac{\partial (\phi^1, \phi^2)}{\partial (x^1, x^2)},
\end{equation}
which is the Jacobian determinant of the mapping from coordinates to the vector field components.This formulation implies that the current $j^\mu$ is supported only at points where $\phi = 0$, identifying these zeros as the critical points of the topological field. The total topological charge $W$ within a domain $\Sigma$ is then given by integrating over the spatial slice \cite{a19, a20}:
\begin{equation}\label{F9}
W = \int_{\Sigma} j^0 \, d^2 x = \sum_{i=1}^n \beta_i \eta_i = \sum_{i=1}^n \overline{\mathcal{\omega}}_i,
\end{equation}
where $\beta_i$ is the Hopf index counting the number of times the vector field loops around the zero point in $\phi$-space, $\eta_i = \text{sign}\bigl( J^0(\phi/x)_{z_i} \bigr) = \pm 1$ determines the orientation, and $\overline{\mathcal{\omega}}_i$ is the winding number associated with each zero point $z_i$. Utilizing the expressions and equations \eqref{function} and \eqref{F1} as groundwork, the explicit form of the free energy reads:
\begin{equation}\label{T1}
\mathcal{F} =\frac{b^2 | \alpha |  \, _2F_1\left(-\frac{1}{2},-\frac{1}{4};\frac{3}{4};-\frac{r_h^4}{b^4}\right)}{2 r_h}-\frac{1}{2} N r_h^{-3 \omega }+\frac{4}{3} \pi  P r_h^3-\frac{\pi  r_h^2}{\tau }+\frac{r_h}{2}.
\end{equation}
The components of the vector field $\phi$, derived from \eqref{F2}, are expressed explicitly as:
\begin{equation}\label{T2}
\begin{split}
&\phi^{r_h} =\frac{b^2 | \alpha |  \left(\, _2F_1\left(-\frac{1}{2},-\frac{1}{4};\frac{3}{4};-\frac{r_h^4}{b^4}\right)-\sqrt{\frac{r_h^4}{b^4}+1}\right)}{2 r_h^2}-\frac{b^2 | \alpha |  \, _2F_1\left(-\frac{1}{2},-\frac{1}{4};\frac{3}{4};-\frac{r_h^4}{b^4}\right)}{2 r_h^2}\\&+\frac{3}{2} N \omega  r_h^{-3 \omega -1}+4 \pi  P r_h^2-\frac{2 \pi  r_h}{\tau }+\frac{1}{2},
\end{split}
\end{equation}
and
\begin{equation}\label{T3}
\phi^{\Theta} = -\frac{\cot(\Theta)}{\sin(\Theta)}.
\end{equation}
Moreover, the parameter $\tau$ takes the form:
\begin{equation}\label{T4}
\tau =\bigg[4 \pi  r_h^{3 \omega +3}\bigg]\times\bigg[-b^2 | \alpha |  \sqrt{\frac{b^4+r_h^4}{b^4}} r_h^{3 \omega }+3 N r_h \omega +8 \pi  P r_h^{3 \omega +4}+r_h^{3 \omega +2}\bigg]^{-1}.
\end{equation}

\begin{figure}[]
\begin{center}
\subfigure[]{
\includegraphics[height=3cm,width=3cm]{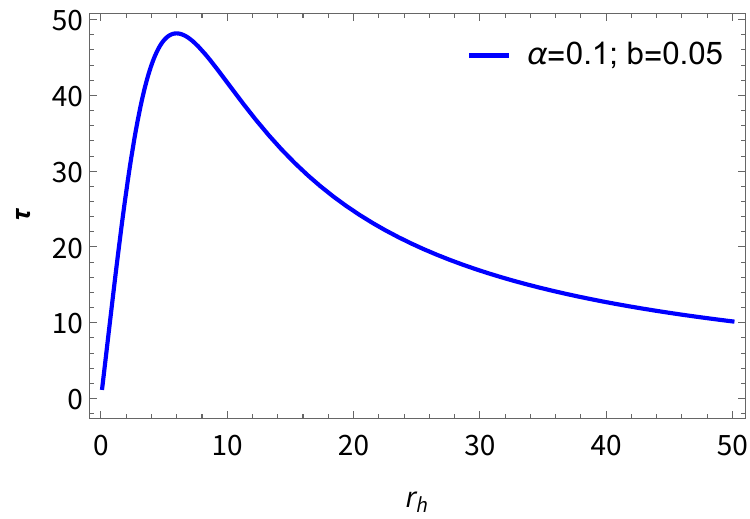}
\label{100a}}\quad\quad
\subfigure[]{
\includegraphics[height=3cm,width=3cm]{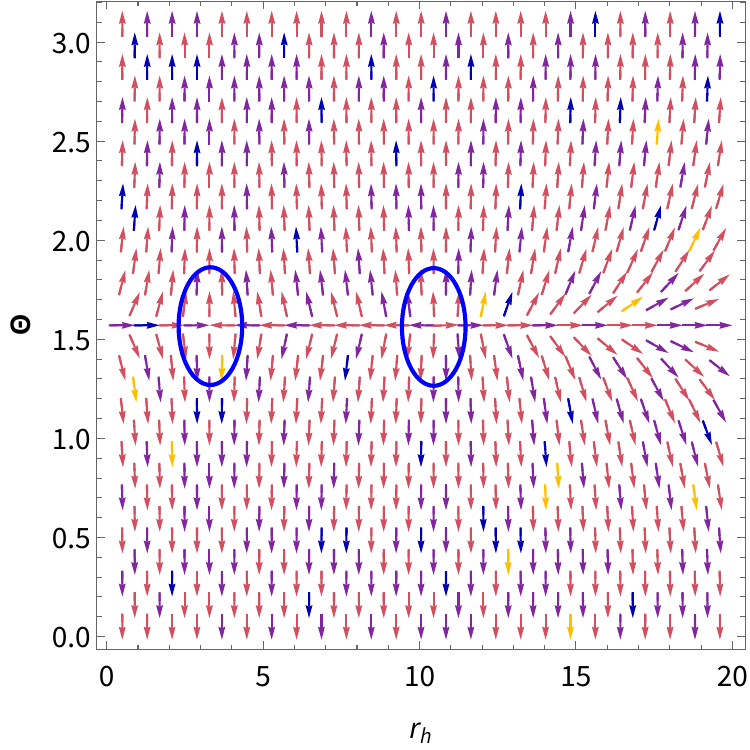}
\label{100b}}\quad\quad
\subfigure[]{
\includegraphics[height=3cm,width=3cm]{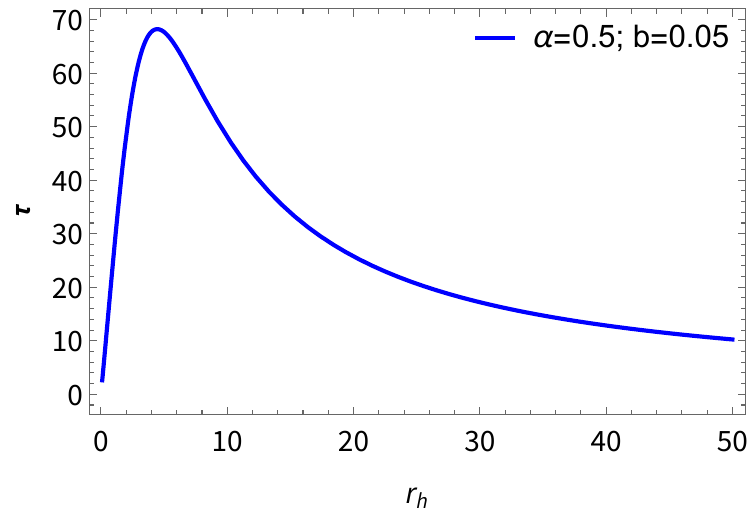}
\label{100c}}\quad\quad
\subfigure[]{
\includegraphics[height=3cm,width=3cm]{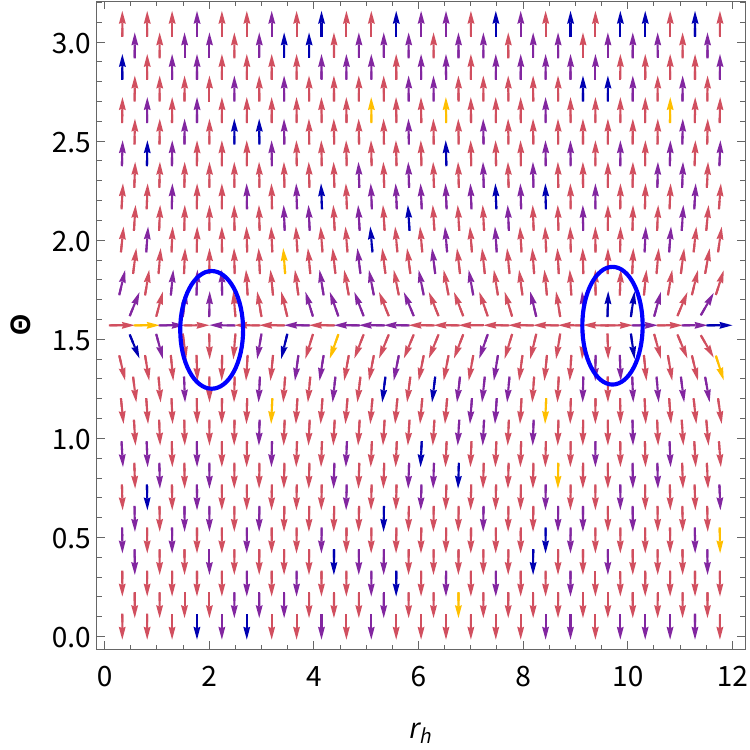}
\label{100d}}\\
\subfigure[]{
\includegraphics[height=3cm,width=3cm]{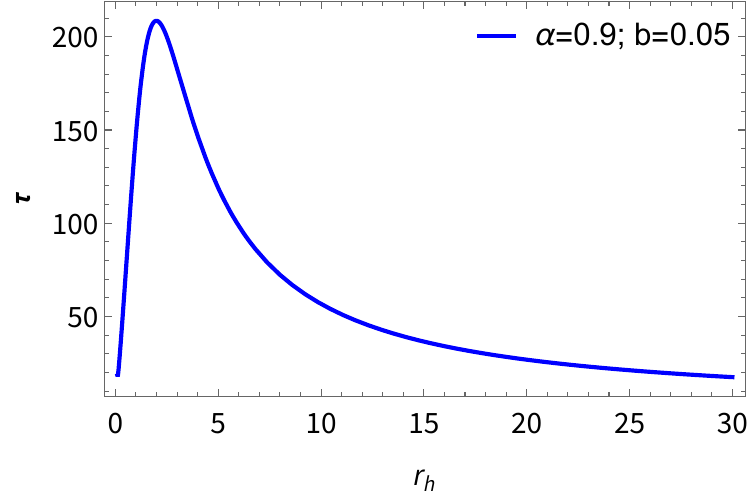}
\label{100e}}\quad\quad
\subfigure[]{
\includegraphics[height=3cm,width=3cm]{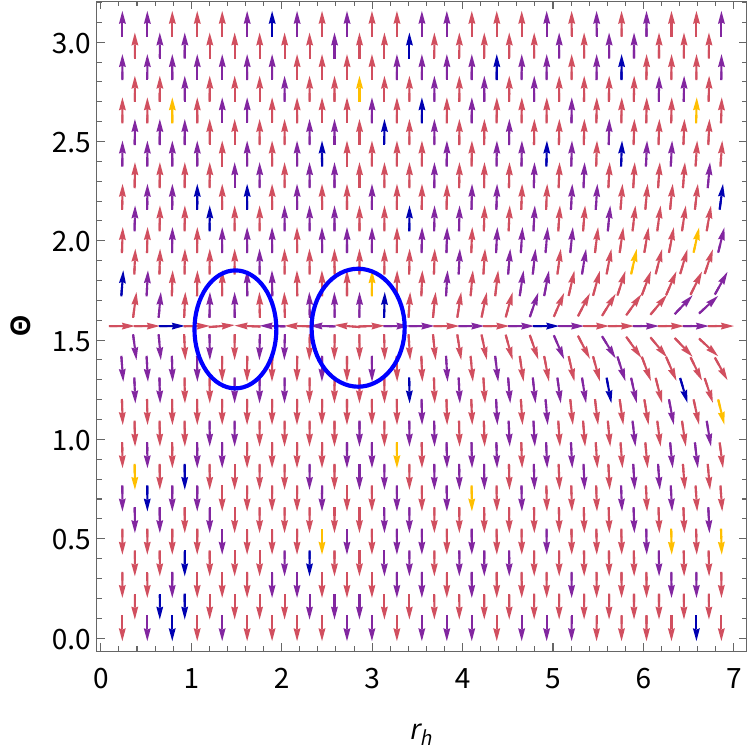}
\label{100f}}\quad\quad
\subfigure[]{
\includegraphics[height=3cm,width=3cm]{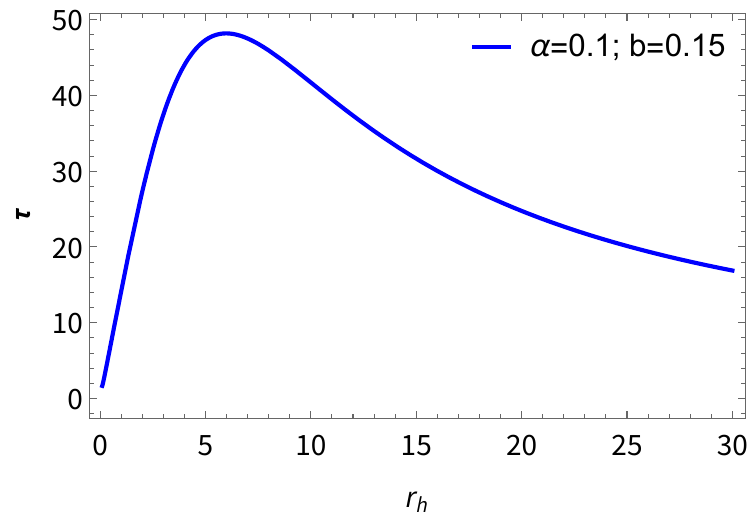}
\label{100g}}\quad\quad
\subfigure[]{
\includegraphics[height=3cm,width=3cm]{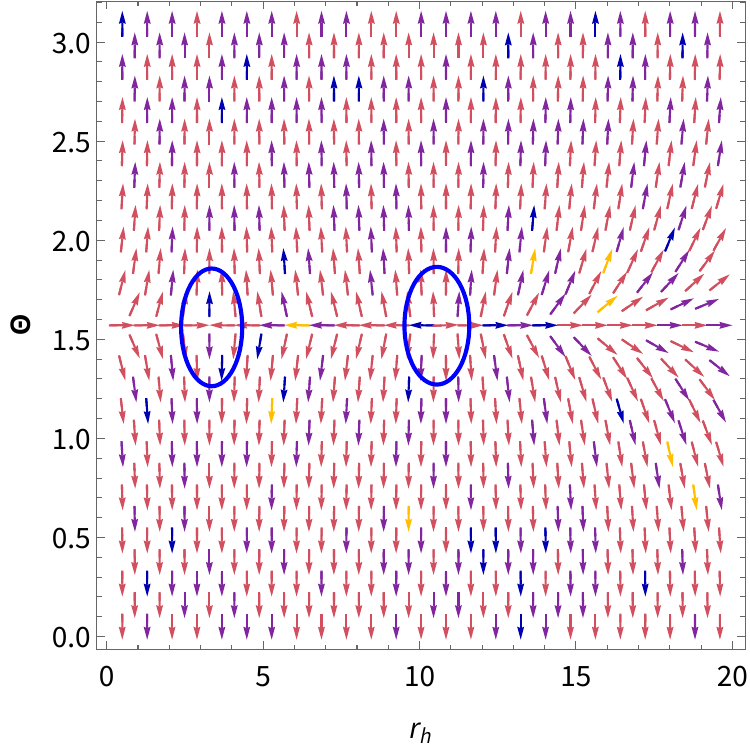}
\label{100h}}\\
\subfigure[]{
\includegraphics[height=3cm,width=3cm]{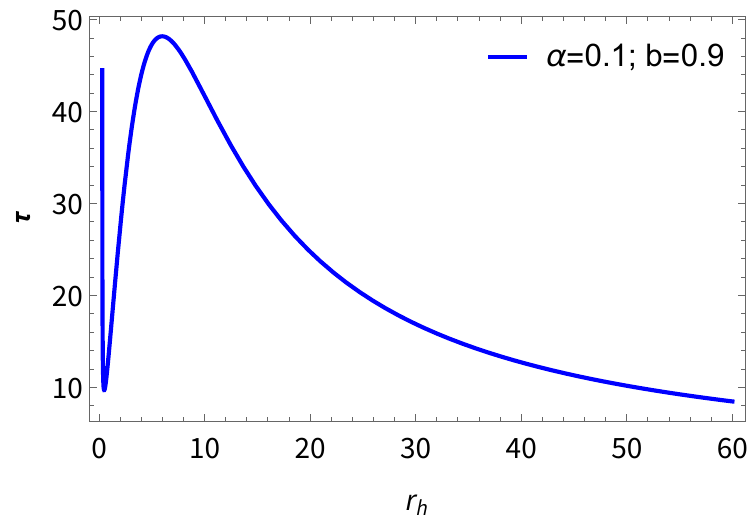}
\label{100i}}\quad\quad
\subfigure[]{
\includegraphics[height=3cm,width=3cm]{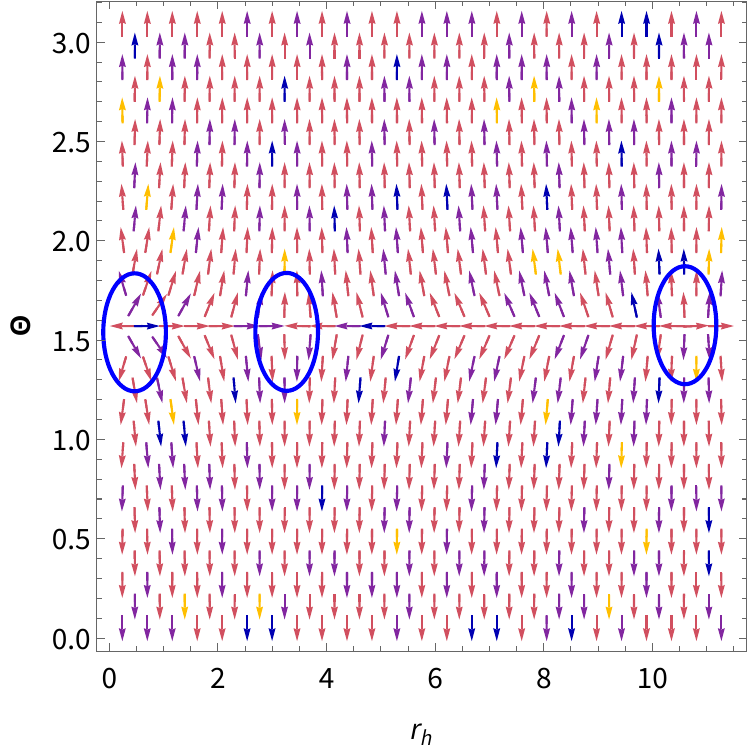}
\label{100j}}\quad\quad
\subfigure[]{
\includegraphics[height=3cm,width=3cm]{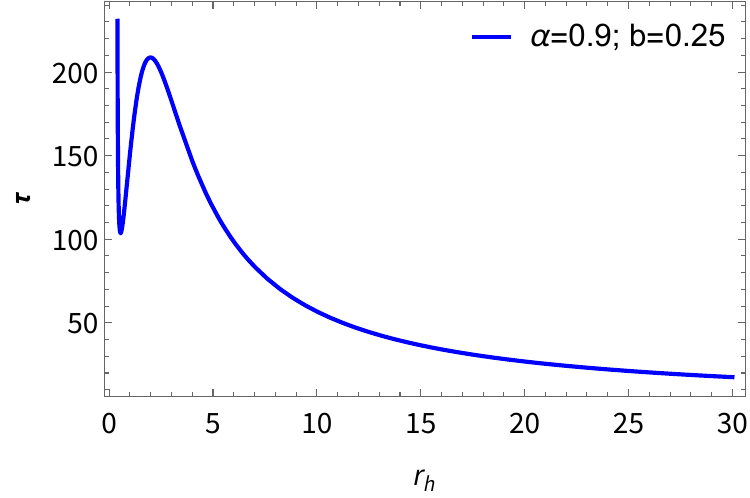}
\label{100k}}\quad\quad
\subfigure[]{
\includegraphics[height=3cm,width=3cm]{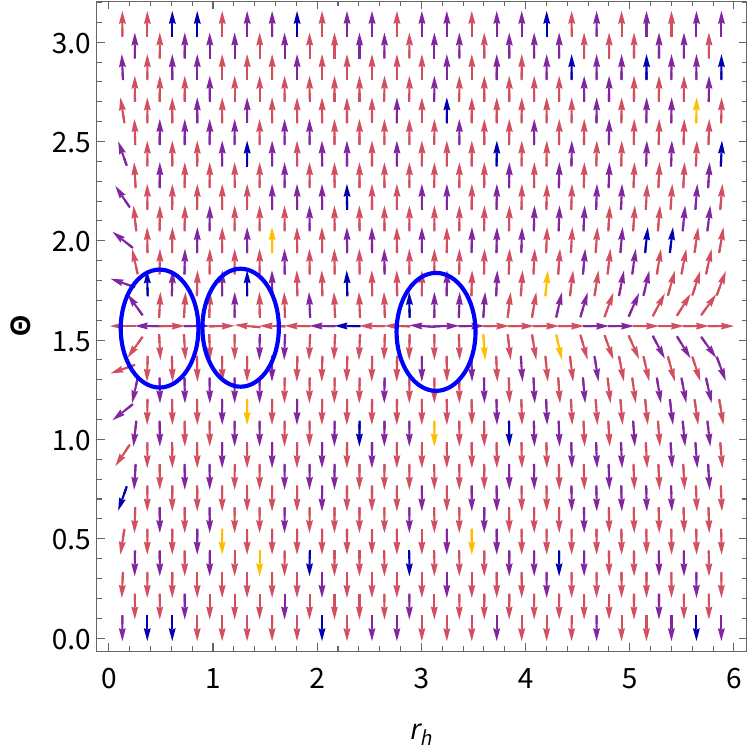}
\label{100l}}\\
\subfigure[]{
\includegraphics[height=3cm,width=3cm]{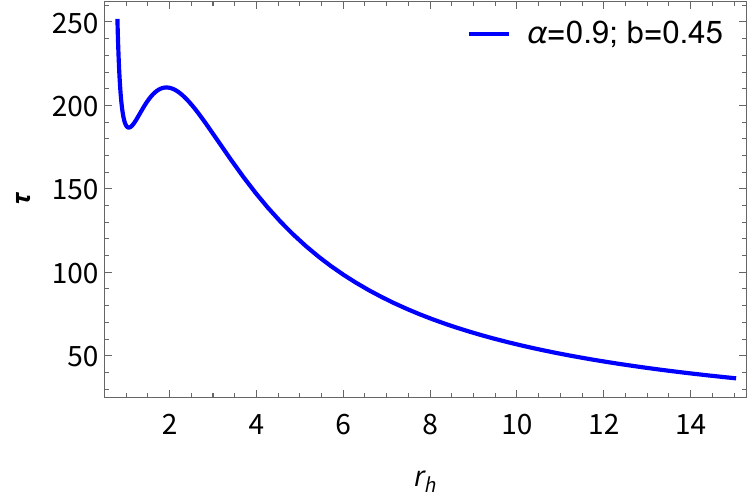}
\label{100m}}\quad\quad
\subfigure[]{
\includegraphics[height=3cm,width=3cm]{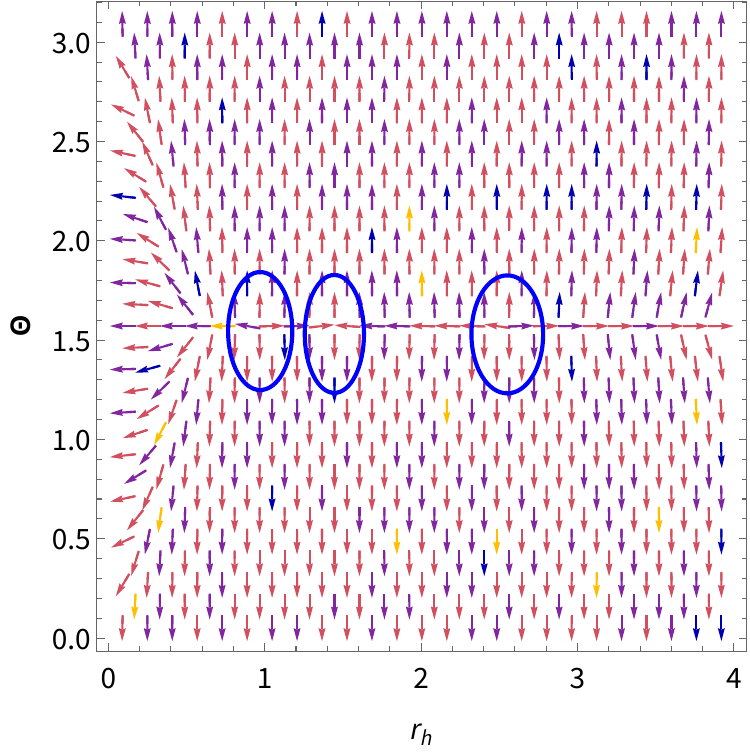}
\label{100n}}\quad\quad
\subfigure[]{
\includegraphics[height=3cm,width=3cm]{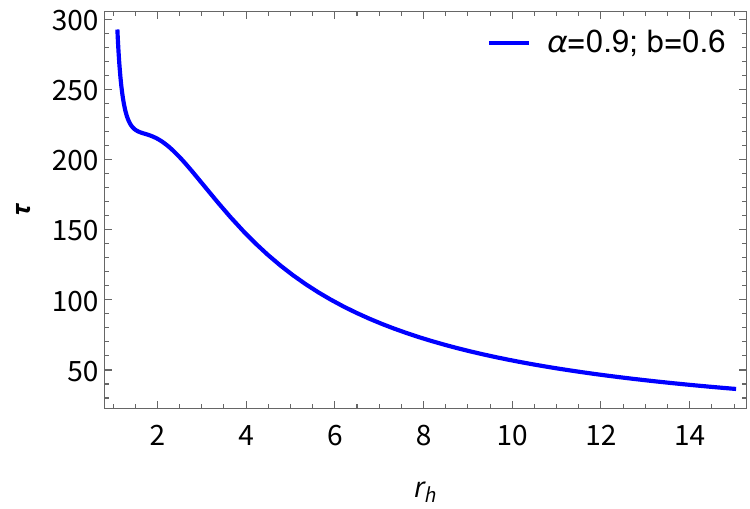}
\label{100o}}\quad\quad
\subfigure[]{
\includegraphics[height=3cm,width=3cm]{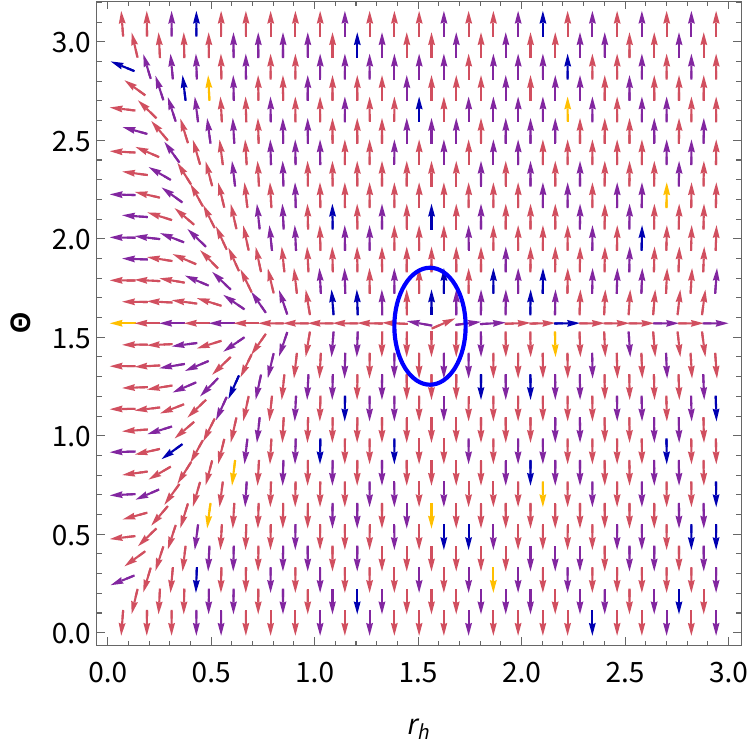}
\label{100p}}\\
\subfigure[]{
\includegraphics[height=3cm,width=3cm]{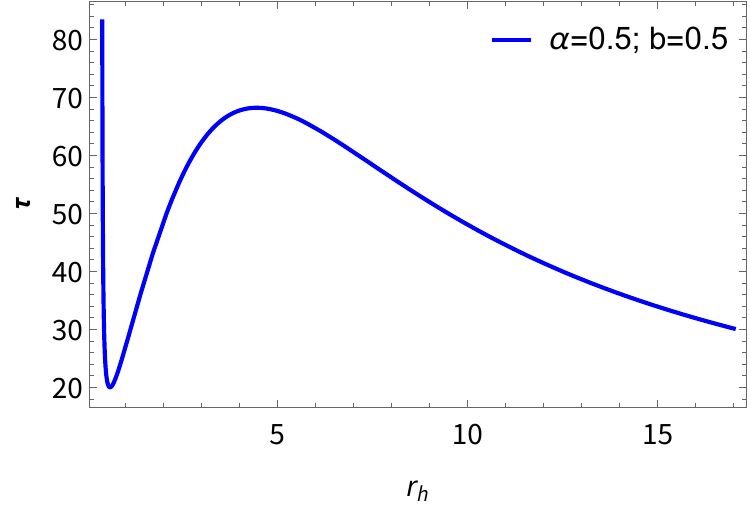}
\label{100q}}\quad\quad
\subfigure[]{
\includegraphics[height=3cm,width=3cm]{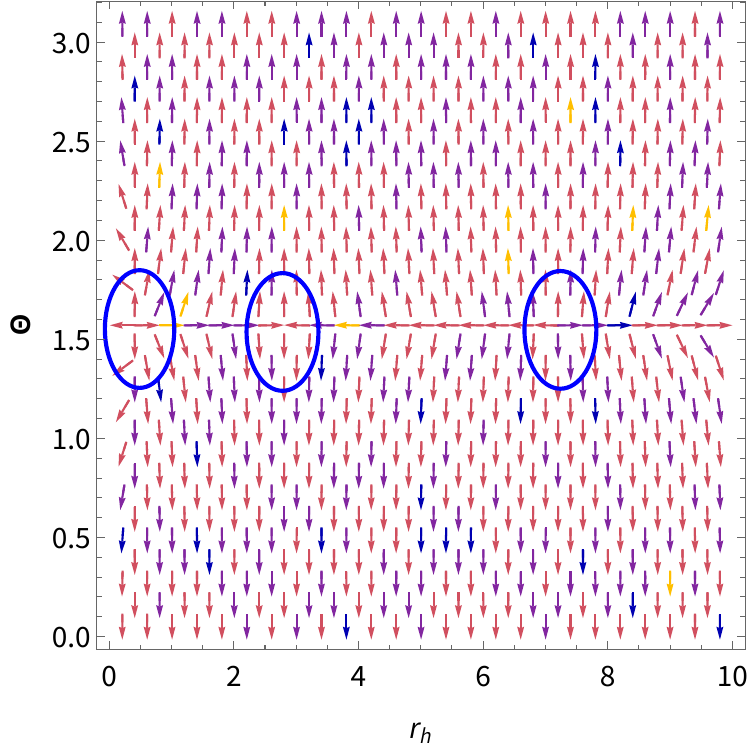}
\label{100r}}
\caption{\small{The behavior of the $\tau\text{-}r_h$ function for the Schwarzschild AdS BHs with a cloud of strings  surrounded by quintessence-like fluid is analyzed within the framework of a normal vector field $n$, which is defined on the $(r_h, \Theta)$-plane. In this analysis, the zero points (ZPs) of the system are identified as the coordinates $(r_h, \Theta)$ at which the vector field vanishes. These zero points correspond to critical locations determined by the underlying structure of the spacetime and are influenced by the free parameters of the model. In particular, the study is conducted for a fixed value of the equation-of-state parameter, $\omega = -\frac{2}{3}$, which typically represents a quintessence-like field in cosmological models. Additionally, the pressure is chosen as $P = 0.001$, and the monopole charge (or topological charge) is set to $N = 0.02$. This specific set of parameters is used to explore the existence, position, and multiplicity of zero points in the $(r_h, \Theta)$-plane, allowing for insights into the geometric and thermodynamic properties of the BH solution under consideration.}}
\label{m1}
\end{center}
\end{figure}

\begin{table}[ht!]
\centering
\setlength{\arrayrulewidth}{0.4mm}
\setlength{\tabcolsep}{2.8pt}
\arrayrulecolor[HTML]{000000}
\begin{tabular}{|>{\centering\arraybackslash}m{2.5cm}|>{\centering\arraybackslash}m{2.5cm}|>{\centering\arraybackslash}m{2.5cm}|>{\centering\arraybackslash}m{3cm}|>{\centering\arraybackslash}m{2.5cm}|}
\hline
\rowcolor[HTML]{9FC5E8}
\textbf{$\omega$} & \textbf{$\alpha$} & \textbf{$b$} & \textbf{$\overline{\omega}$} & \textbf{$W$} \\ \hline
$-\frac{2}{3}$ & 0.1 & 0.05 & $-1,+1$ & $0$ \\ \hline
\rowcolor[HTML]{EAF4FC}
$-\frac{2}{3}$ & 0.5 & 0.05 & $-1,+1$ & $0$ \\ \hline
$-\frac{2}{3}$ & 0.9 & 0.05 & $-1,+1$ & $0$ \\ \hline
\rowcolor[HTML]{EAF4FC}
$-\frac{2}{3}$ & 0.1 & 0.15 & $-1,+1$ & $0$ \\ \hline
$-\frac{2}{3}$ & 0.1 & 0.9 & $+1,-1,+1$ & $+1$ \\ \hline
\rowcolor[HTML]{EAF4FC}
$-\frac{2}{3}$ & 0.9 & 0.25 & $+1,-1,+1$ & $+1$ \\ \hline
$-\frac{2}{3}$ & 0.9 & 0.45 & $+1,-1,+1$ & $+1$ \\ \hline
\rowcolor[HTML]{EAF4FC}
$-\frac{2}{3}$ & 0.9 & 0.6 & $+1$ & $+1$ \\ \hline
$-\frac{2}{3}$ & 0.5 & 0.5 & $+1,-1,+1$ & $+1$ \\ \hline
\end{tabular}
\caption{Topological charges of Schwarzschild AdS BHs with a cloud of strings  surrounded by quintessence-like fluid with respect to free parameters}
\label{P1}
\end{table}

We perform an in-depth investigation of the thermodynamic topology within Schwarzschild AdS BHs with a cloud of strings surrounded by quintessence-like fluid, emphasizing the spatial arrangement of topological charges as presented in Fig.~~(\ref{m1}). The normalized vector field lines shown therein offer a clear visualization of the inherent topological characteristics. Specifically, Fig.~~(\ref{m1}) reveals one, two, and three distinct zeros located at precise coordinates $(r_h, \Theta)$, with corresponding parameter values listed in Table~~(\ref{P1}). These zeros represent localized topological charges encircled by blue contour loops, whose shapes are influenced by changes in free parameters. For instance, when the parameter $b$ is small (e.g., $b=0.05$) and $\alpha$ increases continuously, the system manifests two topological charges, namely $\overline{\omega} = +1$ and $-1$, yielding an overall topological charge $W = 0$, as illustrated in Figs.~~\ref{100b}, \ref{100d}, \ref{100f}, and \ref{100h}. Conversely, upon increasing $b$, a different configuration emerges featuring three charges, $\overline{\omega} = +1, -1, +1$, which sum to a total topological charge of $W = +1$. This is shown in Figs.~~\ref{100j}, \ref{100l}, \ref{100n}, and \ref{100r}. Such behavior persists for values of $\alpha = 0.9$, underscoring the significant influence of this parameter on the BH topology classification. Moreover, simultaneous growth of both $\alpha$ and $b$ leads to the appearance of a single positive topological charge, $\overline{\omega} = +1$, resulting in $W = +1$, as depicted in Fig.~~\ref{100p}. To deepen the understanding of this topological structure, we analyze the free energy as a scalar function over the two-dimensional domain $(r_h, \Theta)$. The corresponding vector field $\phi$ is constructed so that its zeros align with the extremal points of the free energy. The directional circulation of field lines around these zeros, indicative of maxima or minima, enables a rigorous assignment of topological charges to each critical point \cite{a19}.\\

Interpreting the free energy of a BH as a scalar field defined over a two-dimensional parameter space, spanned by the horizon radius $r_h$ and an auxiliary coordinate $\Theta$, allows one to construct a corresponding vector field $\boldsymbol{\phi}$ derived from the gradient of this free energy. The points at which the gradient vanishes coincide exactly with the zeros of $\boldsymbol{\phi}$, representing equilibrium configurations of the BH system.

The behavior of the vector field $\boldsymbol{\phi}$ in the vicinity of these equilibrium points provides crucial insights into the system’s topological properties. Depending on whether the free energy at a critical point corresponds to a local minimum or maximum, the surrounding vector field lines exhibit characteristic rotational patterns. These can be quantified by assigning a winding number, or topological charge, to each zero of $\boldsymbol{\phi}$. This charge effectively measures the number of times the vector field encircles the zero, serving as a topological invariant that captures the nature and stability of the equilibrium.

Earlier studies on classical BHs, such as the Schwarzschild, Reissner-Nordström, and Anti-de Sitter (AdS) Reissner-Nordström solutions, have identified distinct topological charges associated with each configuration. Specifically, the Schwarzschild BH possesses a total topological charge $W = -1$. The Reissner-Nordström BH, which incorporates electric charge, carries a neutral charge $W = 0$, while the AdS-Reissner-Nordström BH shows a positive charge $W = +1$, reflecting differences in thermodynamic phases and stability \cite{a19}. These charges encode fundamental distinctions in the underlying phase structure of these BH types.

Serving as canonical examples in gravitational physics, these topological classifications provide a baseline for categorizing a wider class of BH solutions. As summarized in Table~~(\ref{P1}), BHs with $W = +1$ tend to mimic the thermodynamic and phase behavior characteristic of AdS-Reissner-Nordström solutions, whereas those with $W = 0$ exhibit features more closely aligned with Reissner-Nordström BHs \cite{a19}.

In our current framework (Schwarzschild-AdS black holes with cloud of strings and quintessence), we examine how the topology of BH solutions evolves as the system parameters vary. Adjusting these free parameters modifies the distribution of zeros in $\boldsymbol{\phi}$, thereby altering the topological charge assignments and consequently the thermodynamic stability and phase structure. Notably, configurations with total topological charge $W = +1$ behave analogously to AdS-Reissner-Nordström BHs, while those with $W = 0$ align with Reissner-Nordström-like thermodynamics.

Importantly, the parameter space itself plays a decisive role in determining the pattern of topological charges. Modifications in parameters, interpreted as effects arising from quantum corrections or geometric deformations, can change the number and type of zeros of the vector field $\boldsymbol{\phi}$, which in turn influence the possible phase transitions and stability regimes of the BH system.

The parameter sensitivity of the total topological charge, as demonstrated in the figures associated with Table~~(\ref{P1}), underscores the critical influence of model parameters on BH thermodynamics. This intricate relationship between free energy topology and parameter variation presents a powerful framework for classifying BH solutions, anticipating their phase behavior, and probing the impact of modifications to classical gravity. For conclusive remarks, we base our analysis on the total topological charge determined at each parameter setting, thoroughly exploring the implications of these variations on the system’s equilibrium and phase structure.

\section{Scalar Field Perturbations and Quasinormal Mode Spectrum}\label{sec:5}

Studying scalar field perturbations provides insights into the linear stability of the BH solution and yields quasinormal modes (QNMs), which encode the ringdown signatures of BHs in gravitational wave detections and thermalization processes in the dual field theory via AdS/CFT. To explore the dynamical stability and response of the Schwarzschild-AdS BH with a cloud of strings (CoS) and a surrounding quintessence-like fluid (QF), we consider massless scalar field perturbations governed by the Klein-Gordon equation in curved spacetime:
\begin{equation}
	\Box \Phi = \frac{1}{\sqrt{-g}} \partial_\mu \left( \sqrt{-g} g^{\mu\nu} \partial_\nu \Phi \right) = 0.
\end{equation}
Assuming a separable ansatz of the form,
\begin{equation}
	\Phi(t, r, \theta, \phi) = e^{-i \omega t} Y_{\ell m}(\theta, \phi) \frac{\psi(r)}{r},
\end{equation}
and applying this to the metric in Eq.~(\ref{final}), we arrive at the Schrödinger-like master equation for the radial part:
\begin{equation}\label{Slike}
	\frac{d^2 \psi}{dr_*^2} + \left[ \omega^2 - V_\text{scalar}(r) \right] \psi = 0,
\end{equation}
where $r_*$ is the tortoise coordinate defined as $dr_* = dr / f(r)$.
It is important to note that the expression for the effective potential in asymptotically AdS spacetime assumes a standard Schwarzschild-like form for the metric. In our case, the presence of the cloud of strings (CoS) and quintessence-like fluid (QF) modifies the structure of the metric function $ f(r) $, potentially altering the behavior of the potential at large $ r $. We proceed with the following expression for the effective potential under the assumption that the angular and radial parts remain separable and well-behaved:
\begin{align}\label{effV}
	V_{\text{scalar}}(r) &= f(r)\left[\frac{\ell(\ell+1)}{r^2} + \frac{f'(r)}{r}\right]\nonumber\\
    &=\left(1 - \frac{2\,M}{r} + \frac{\lvert \alpha\rvert\, b^2}{r^2}\,{}_2F_1\left(-\frac{1}{2},-\frac{1}{4},\frac{3}{4},-\frac{r^4}{b^4}\right)-\frac{\mathrm{N}}{r^{3\,w+1}}+\frac{r^2}{\ell^2_p}\right)\times\nonumber\\
    &\left(\frac{\ell(\ell+1)}{r^2}+\frac{2\,M}{r^3}
- \frac{2\,|\alpha|\, b^2}{r^4}\, {}_2F_1\left(-\tfrac{1}{2}, -\tfrac{1}{4}, \tfrac{3}{4}, -\tfrac{r^4}{b^4} \right)
- \frac{2|\,\alpha|}{3\, b^2}\, {}_2F_1\left(\tfrac{1}{2}, \tfrac{3}{4}, \tfrac{7}{4}, -\tfrac{r^4}{b^4} \right)
+ \frac{\mathrm{N}\,(3\,w+1)}{r^{3\,w+3}}
+ \frac{2}{\ell_p^2}\right) 
\end{align}
This form holds under the assumption that the dominant contribution at large $ r $ comes from the AdS curvature term.
The shape of $V_\text{scalar}(r)$ is influenced by the CoS and QF parameters $(\alpha, b, \mathrm{N}, w)$ as well as the AdS curvature radius $\ell_p$. 
The effective potential $ V_{\text{scalar}}(r) $ for scalar perturbations, given in Eq.~(\ref{effV}), plays a crucial role in determining the dynamical response of the BH spacetime under linear perturbations. Its form encapsulates the influence of the background geometry, specifically through the metric function $ f(r) $, which in turn depends on the presence of the cloud of strings and the surrounding quintessence-like fluid. The structure of the potential reveals key information about wave propagation: its height and width govern the scattering behavior of scalar waves, while the location of its peak corresponds to the region of strongest gravitational trapping.

\begin{figure}[ht!]
    \centering
    \includegraphics[width=0.45\linewidth]{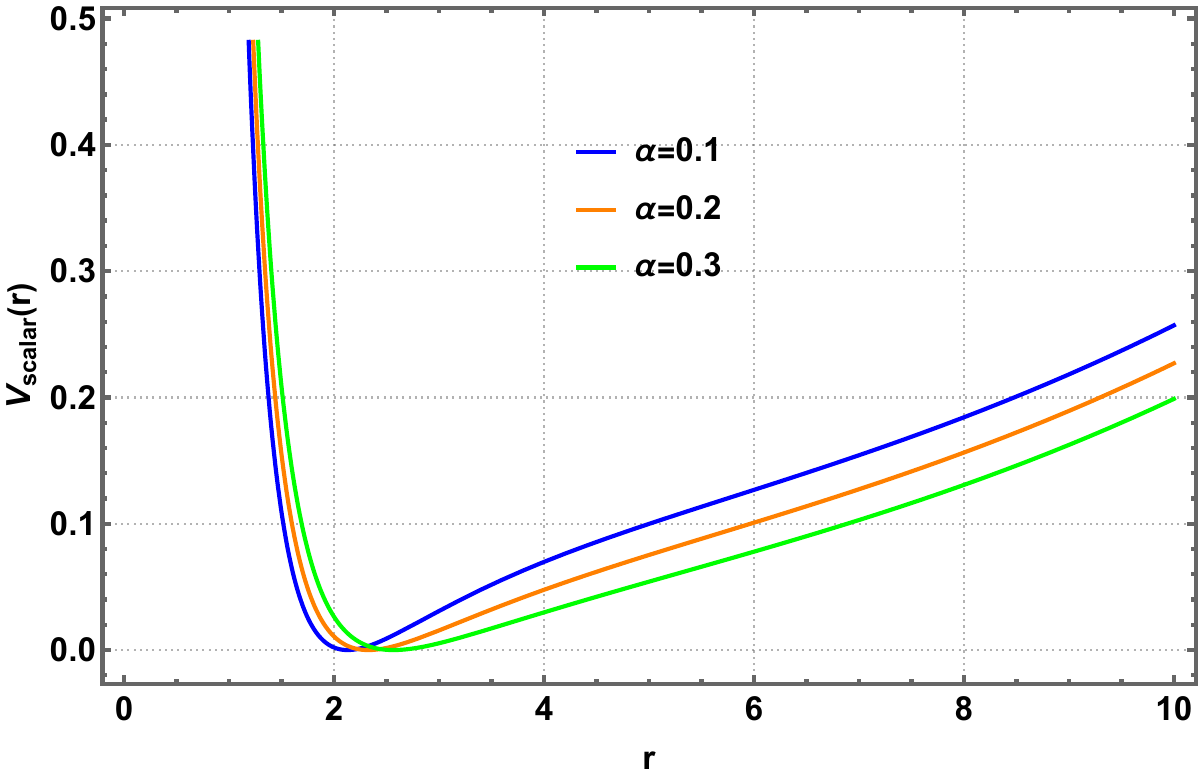}\quad\quad\quad
    \includegraphics[width=0.45\linewidth]{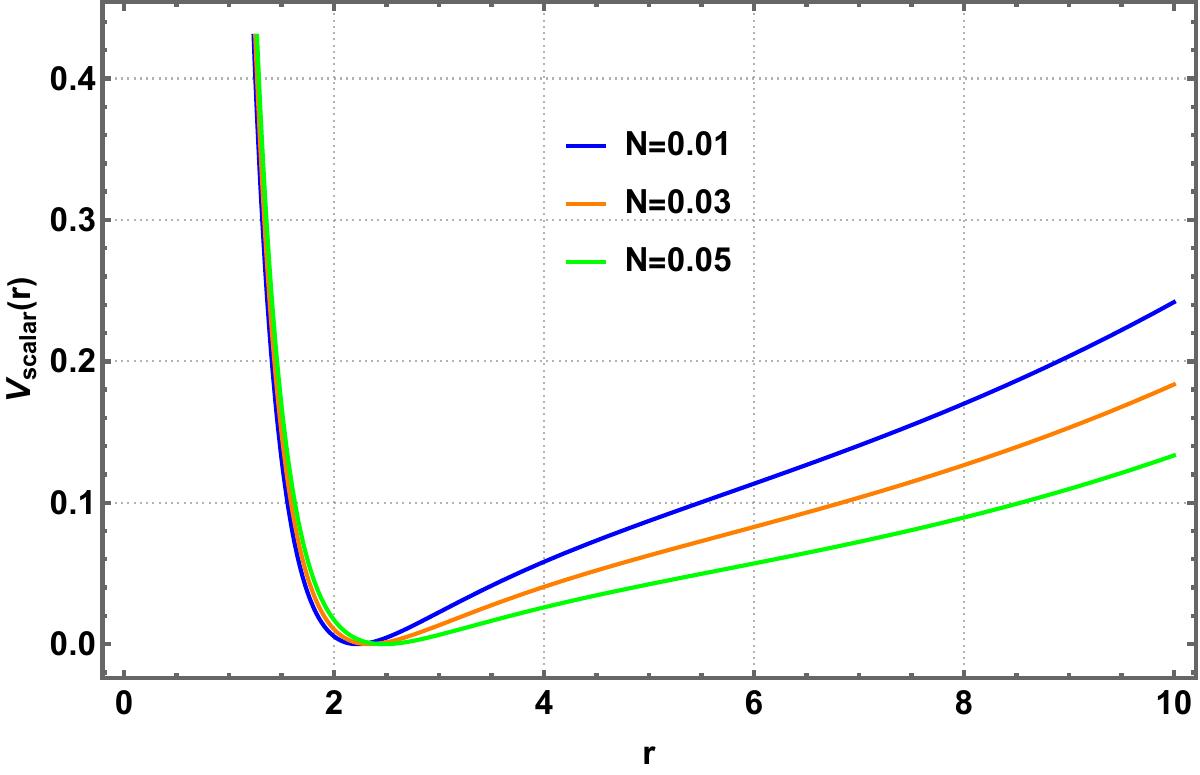}\\
    (a) $\mathrm{N}=0.01$  \hspace{6cm} (b) $\alpha=0.15$
    \caption{\footnotesize Behavior of the scalar perturbative potential $V_\text{scalar}(r)$ for $\ell=0$ modes by varying the parameters $\alpha$ and $\mathrm{N}$. Here, we set $M=1, \ell_p=10, w=-2/3, b=0.2$.}
    \label{fig:perturbative}
\end{figure}

\begin{figure}[h!]
	\centering
	\includegraphics[width=0.5\linewidth]{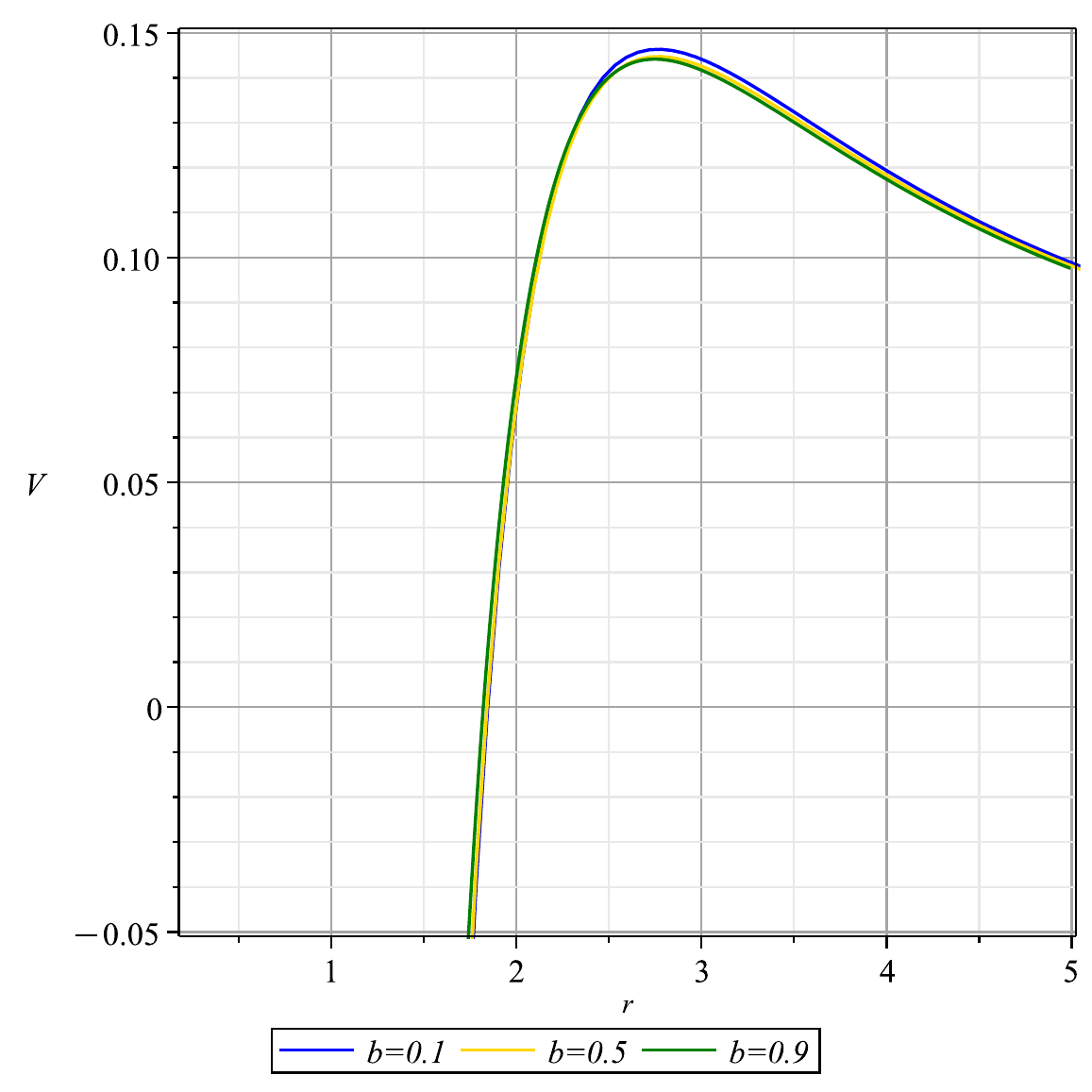}\includegraphics[width=0.5\linewidth]{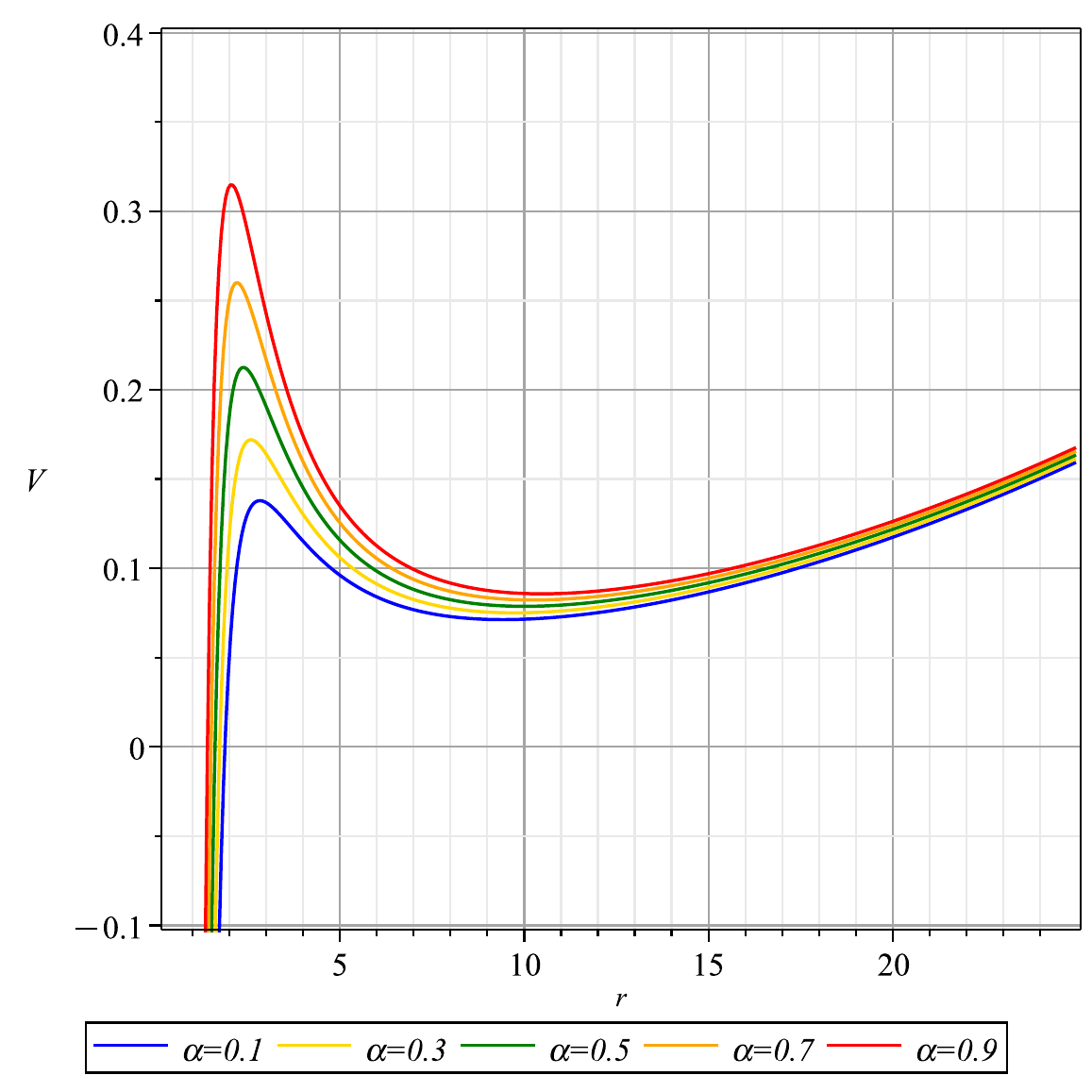}
	\caption{Effective potential $V_{\mathrm{eff}}(r)$ for scalar field perturbations with $\ell=1$ and fixed values $M=1$, $N=0.01$, $w=-2/3$, $\ell_p=10$. Each curve corresponds to a different combination of the string cloud parameters $(\alpha, b)$. Left: for fixed $\alpha$ and varying $b$. Right: for varying  $\alpha$ while fixed $b$.}
	\label{fig:Veff_multi}
\end{figure}

Fig.~\ref{fig:perturbative} illustrates the behavior of the scalar perturbative potential $V_\text{scalar}$ for $\ell=0$ mode by varying the string parameter $\alpha$ and normalization constant $\mathrm{N}$ of QF, while keeping other parameters fixed.
Fig. \ref{fig:Veff_multi} depicts the radial profile of the effective potential $V_{\mathrm{scalar}}(r)$ for scalar field perturbations in the Schwarzschild-AdS BH background. The potential is shown for various values of the string cloud parameters $\alpha$ and $b$, while keeping the BH mass $M=1$, state parameter $w=-2/3$, normalization constant $\mathrm{N}=0.01$, AdS radius $\ell_p = 10$, and scalar field angular index $\ell=1$ fixed. The horizontal axis corresponds to the radial coordinate $r$, extending from near the event horizon outwards, while the vertical axis shows the value of the effective potential. The plotted curves correspond to different parameter choices (e.g., coupling constants or BH parameters) and illustrate how the potential barrier changes with these settings. The figure shows that $V_{\mathrm{scalar}}(r)$ vanishes at the event horizon, as expected from the regularity of the field equation there. Away from the horizon, the potential rises to a local maximum, forming a barrier that governs the scattering and quasinormal mode spectrum of the scalar field. The height and position of this peak depend on the chosen parameters: stronger coupling or smaller BH radius generally increases the barrier height and shifts the maximum closer to the horizon. At large $r$, the potential asymptotically increases due to the confining nature of the Anti-de Sitter spacetime, which traps perturbations and leads to a discrete spectrum. Physically, this barrier structure implies that scalar waves experience partial reflection and transmission, with the AdS boundary acting as a reflective wall. The variation in peak height influences the decay rates and oscillation frequencies of quasinormal modes, while the asymptotic growth ensures the stability of perturbations in this setup.

To determine the quasinormal mode (QNM) spectrum, we consider the appropriate physical boundary conditions associated with asymptotically AdS BH spacetimes. Specifically, the scalar perturbation must represent purely ingoing waves at the BH event horizon ($r \to r_h$), ensuring that no information escapes from within. At spatial infinity ($r \to \infty$), the field is required to vanish, reflecting the confining nature of the AdS boundary. These conditions guarantee a well-posed eigenvalue problem for the QNM frequencies and are consistent with the reflective behavior typical of AdS geometries.

To compute the quasinormal mode frequencies, we use the sixth-order WKB approximation method \cite{konoplya2003quasinormal}. While commonly used, it is important to emphasize that the WKB approximation has known limitations in asymptotically AdS spacetimes. This is primarily because the WKB method is ideally suited to potentials that vanish at both spatial infinity and the event horizon, with localized barrier structure. However, in AdS backgrounds, the potential diverges at spatial infinity due to the reflective boundary, which introduces nontrivial modifications to wave propagation.
Therefore, while the WKB method gives qualitatively reasonable estimates for low-lying modes, more accurate computations would require methods adapted to AdS geometries, such as Leaver’s continued fraction method or the Horowitz-Hubeny technique. Therefore we have,
\begin{equation}
	\omega^2 = V_0 + \sqrt{-2 V_0''} \, \Lambda(n) - i \left( n + \frac{1}{2} \right) \sqrt{-2 V_0''} \left[ 1 + \Omega(n) \right],
\end{equation}
where $V_0$ is the potential maximum, $V_0''$ its second derivative at the peak, and $\Lambda(n)$, $\Omega(n)$ are higher-order correction terms.
For illustrative purposes, we numerically compute QNMs for the scalar field with $\ell = 1$, $n = 0$ (fundamental mode), for varying $\alpha$, keeping $M = 1$, $\mathrm{N} = 0.01$, $w = -2/3$, and $\ell_p = 10$ fixed.
The quantities $ V_0 $ and $ V_0'' $, representing the peak and curvature of the effective potential respectively, are computed numerically for each parameter set. We ensure that the potential admits a single maximum for the considered values, which validates the application of the WKB approximation to first order. Although we do not explicitly present the computed values of $ V_0 $ and its derivatives, these were obtained by evaluating $ V_{\text{eff}}(r) $ numerically using standard finite-difference methods.

\begin{table}[ht!]
	\centering
	\begin{tabular}{|c|c|c|c|}
		\hline
		$\alpha$ & $b$ & $\text{Re}(\omega)$ & $\text{Im}(\omega)$ \\
		\hline
		0.1 & 0.2 & 0.456 & -0.093 \\
		0.2 & 0.2 & 0.432 & -0.096 \\
		0.3 & 0.2 & 0.401 & -0.101 \\
		0.4 & 0.2 & 0.364 & -0.108 \\
		\hline
	\end{tabular}
	\caption{QNMs frequencies of the scalar field ($\ell = 1, n = 0$) for selected values of $\alpha$ with $b=0.2$, $\mathrm{N}=0.01$, and $\ell_p = 10$.}
	\label{tab:4}
\end{table}

From Tab. \ref{tab:4}, we observe that increasing the CoS parameter $\alpha$ results in lower oscillation frequency $\text{Re}(\omega)$ and larger damping rate $\lvert \text{Im}(\omega) \rvert$, indicating stronger dissipation due to enhanced gravitational interaction near the BH.
This behavior confirms the dynamical stability of the system under scalar perturbations, while also providing a potential observational signature of string clouds and quintessence fields via gravitational wave echoes in AdS/CFT setups.

To further explore the role of the surrounding quintessence-like fluid on the dynamical response of the BH, we now investigate how the quasinormal mode (QNM) spectrum depends on variations of the state parameter $ w $ in the allowed range $ -1 < w < -\frac{2}{3} $. Physically, this parameter governs the equation of state of the dark energy component, with lower values of $ w $ corresponding to stronger negative pressures (see Tab. \ref{tab:qnm_w_dependence}). By analyzing scalar field perturbations for different choices of $ w $, we can gain deeper insight into how the fluid's nature modifies the effective potential and thus alters the decay properties of perturbations.

Numerical results show that decreasing $ w $ (i.e., making it more negative) tends to increase the height and steepness of the effective potential barrier. This results in stronger confinement of the scalar field near the BH, enhancing the curvature effects and leading to more rapidly damped modes. Consequently, the imaginary part $ \text{Im}(\omega) $ of the QNM frequency becomes more negative, indicating faster decay of perturbations. On the other hand, the real part $ \text{Re}(\omega) $, which determines the oscillation frequency, shows a mild decrease as $ w $ becomes more negative. This behavior reflects the influence of the quintessence field on the effective gravitational potential experienced by the scalar wave.

These findings suggest that the quintessence state parameter plays a non-trivial role in shaping the QNM spectrum, especially in AdS backgrounds where boundary conditions enforce reflective behavior. In the context of the AdS/CFT correspondence, such variations in damping rates could be interpreted as changes in the thermalization timescale of the dual field theory. Therefore, by tuning $ w $, one may in principle influence the relaxation dynamics of perturbations, offering a novel observational window into the nature of dark energy surrounding BHs.

\begin{table}[ht!]
	\centering
	\caption{Quasinormal mode frequencies of scalar field perturbations ($ \ell = 1, n = 0 $) for varying quintessence state parameters $ w $. Parameters are fixed at $ \alpha = 0.2, b = 0.2, M = 1, \mathrm{N} = 0.01, \ell_p = 10 $.}
	\begin{tabular}{|c|c|c|}
		\toprule
		$w$ & $\text{Re}(\omega)$ & $\text{Im}(\omega)$ \\
		\midrule
		$-0.67$ & 0.435 & -0.094 \\
		$-0.70$ & 0.429 & -0.096 \\
		$-0.75$ & 0.422 & -0.099 \\
		$-0.80$ & 0.417 & -0.102 \\
		$-0.90$ & 0.409 & -0.106 \\
		$-0.99$ & 0.403 & -0.110 \\
		\bottomrule
	\end{tabular}
	\label{tab:qnm_w_dependence}
\end{table}

\begin{figure}[ht!]
	\centering
	\includegraphics[width=0.7\textwidth]{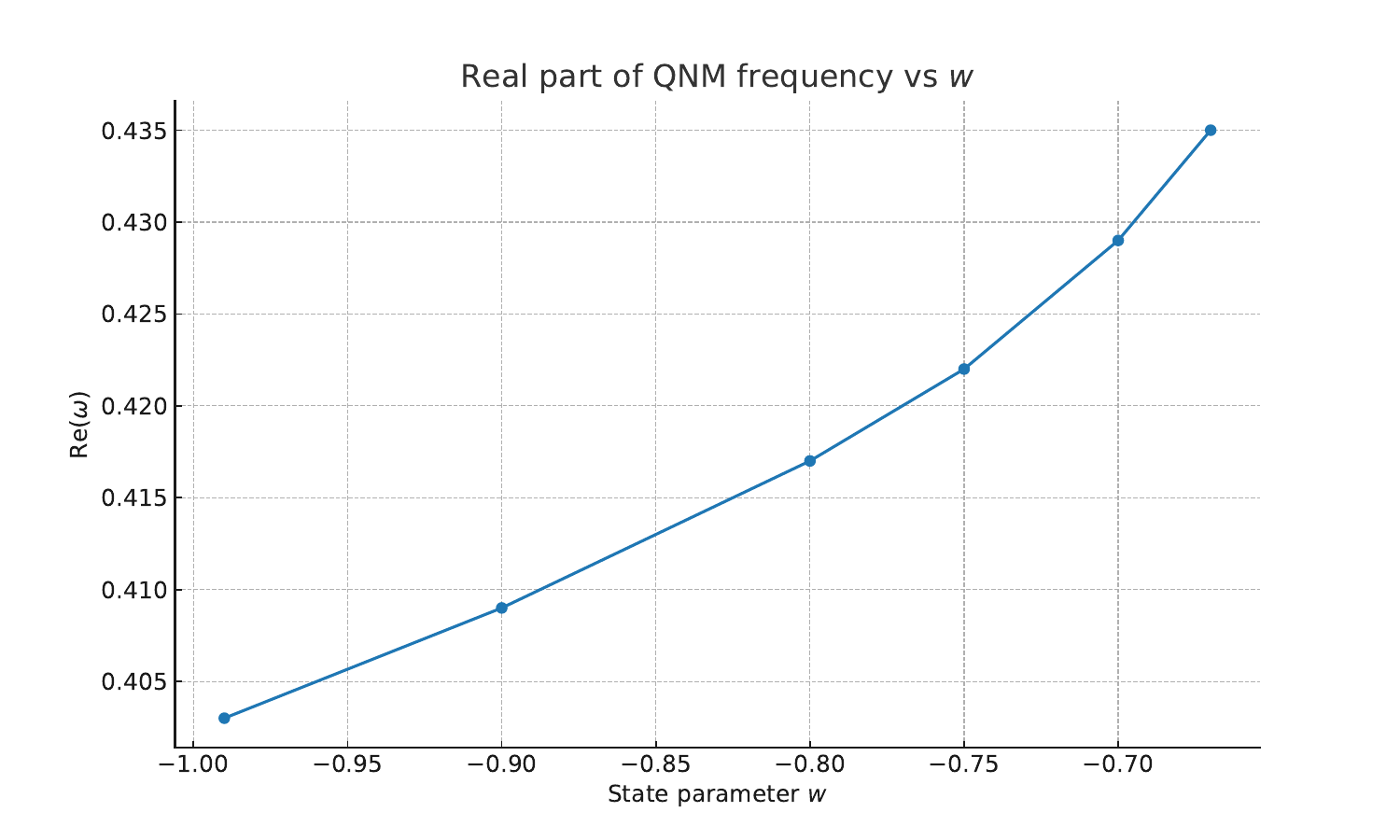}
	\caption{Real part of the QNM frequency as a function of the quintessence parameter $ w $ for scalar field perturbations.}
	\label{fig:re_qnm_w}
\end{figure}

\begin{figure}[ht!]
	\centering
	\includegraphics[width=0.7\textwidth]{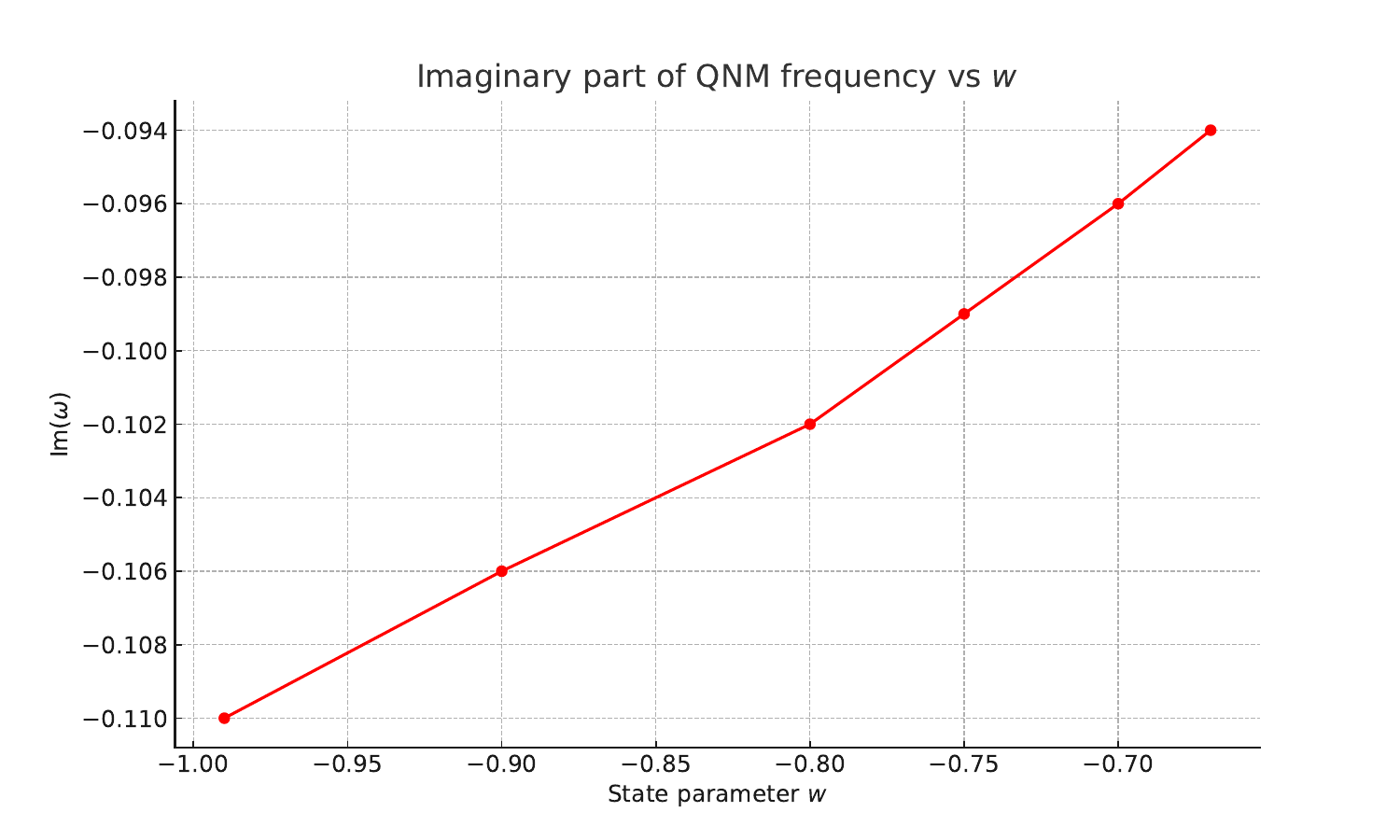}
	\caption{Imaginary part of the QNM frequency as a function of the quintessence parameter $ w $. Increasing negative pressure (lower $ w $) leads to faster decay of perturbations.}
	\label{fig:im_qnm_w}
\end{figure}

Figures~\ref{fig:re_qnm_w} and~\ref{fig:im_qnm_w} together illustrate the impact of the quintessence state parameter $ w $ on the real and imaginary parts of the quasinormal mode (QNM) frequencies. Both quantities exhibit a decreasing trend as $ w $ becomes more negative, reflecting a consistent influence of the surrounding quintessence-like fluid on the scalar perturbations.

As shown in Fig.~\ref{fig:re_qnm_w}, the real part $ \text{Re}(\omega) $, which determines the oscillation frequency of the perturbation, gradually decreases with decreasing $ w $. This indicates that the presence of a more dominant quintessence field slows down the oscillatory dynamics of the scalar field.

Fig.~\ref{fig:im_qnm_w} shows a similar monotonic decrease in the imaginary part $ \text{Im}(\omega) $, which is negative due to damping. Importantly, the absolute value $ |\text{Im}(\omega)| $ increases as $ w $ decreases, signifying that the perturbations decay more rapidly in stronger quintessence environments. This enhanced damping is consistent with the gravitational potential becoming more confining, leading to faster dissipation of scalar waves. Therefore, the plots reveal that more negative values of $ w $ lead to both slower oscillations and faster damping of perturbations, a signature of increased curvature effects due to the quintessence field. In the holographic context, this behavior can be interpreted as shorter thermal relaxation times in the dual conformal field theory.

In the context of the AdS/CFT correspondence, the analysis of scalar field perturbations and their associated quasinormal modes (QNMs) carries significant implications for the thermal behavior of the dual conformal field theory. Within this framework, the frequencies of QNMs are interpreted as the poles of the retarded Green's functions of corresponding operators in the boundary theory. Specifically, the imaginary part of the QNM frequencies governs the decay rate of perturbations in the bulk and is directly related to the relaxation timescale in the dual CFT. Hence, a larger absolute value of $\text{Im}(\omega)$ implies faster thermalization of the boundary field theory after a perturbation.
Our results show that the presence of a cloud of strings (CoS) and a quintessence-like fluid (QF) significantly influences the damping rates of scalar perturbations. For instance, increasing the CoS parameter $\alpha$ leads to a steeper effective potential barrier and correspondingly increases the imaginary component of the QNM frequency. This behavior suggests that the dual CFT thermalizes more rapidly when the gravitational background is modified by stronger string cloud effects. Conversely, modifying the parameter $b$ (related to the spatial distribution of the CoS) leads to a broader and lower potential, implying a more gradual decay of perturbations and, hence, slower boundary thermalization.
These findings point to a deeper interplay between the bulk matter configuration and the holographic response of the boundary theory. In particular, the role of exotic matter fields such as the CoS and QF may leave imprints on the late-time dynamics of strongly coupled field theories through their impact on the BH's QNM spectrum. Such insights could be relevant in holographic models of non-equilibrium phenomena, where external fields or matter components play a role in the equilibration process.

While our study demonstrates how the cloud of strings and quintessence fluid affect the scalar perturbation spectrum via the WKB approximation, we emphasize that a more rigorous analysis would benefit from methods tailored to the asymptotically AdS background. In particular, the Horowitz-Hubeny approach or Leaver's method can accurately capture the reflective nature of the AdS boundary and improve the reliability of quasinormal mode spectra, especially for higher overtones or near-extremal regimes. Future work may include implementing these techniques for a comprehensive stability analysis.

\section{Conclusions} \label{sec:6}

In this study, we conducted a comprehensive investigation of geodesic motion around a Schwarzschild-AdS BH coupled with a cloud of strings and surrounded by a quintessence-like fluid. We derived the effective potential governing test particle dynamics and analyzed key features such as effective radial forces, photon trajectories, the photon sphere, and the BH (BH) shadow. Our results demonstrate that both CoS and QF significantly modify these optical features, leading to observable deviations from the standard AdS BH solution. We further examined the topology of photon spheres-critical surfaces where light can orbit the BH and their role in stability analysis. Each photon sphere corresponds to a zero of the vector field associated with the effective geometry and carries a quantized topological charge of either $0$ or $-1$, depending on the winding number of the vector field. Our parametric analysis reveals that the total topological charge associated with photon spheres remains invariant at $-1$ across a broad range of physical parameters. This invariance reflects deep geometric constraints on photon orbits and underscores their utility as robust indicators of BH stability. By analyzing the structure of photon spheres and the associated effective potentials, we explored how parameters such as $\alpha$ and $b$ influence their topological characteristics. Notably, the emergence of unstable photon spheres serves as a signature of phase transitions and dynamical instabilities within the BH system, offering a powerful diagnostic tool in BH thermodynamics. Additionally, we investigated the motion of massive test particles, calculating the specific energy and angular momentum required for stable circular orbits. We also analyzed the innermost stable circular orbit and demonstrated how CoS and QF parameters influence its location, further highlighting the impact of these fields on particle dynamics in curved space-time.

We investigated the thermodynamic topology of BH (BH) systems by analyzing the spatial distribution and classification of topological charges associated with the zeros of a constructed vector field. The analysis, visualized through normalized vector field lines, reveals that the BH system can exhibit configurations containing one, two, or three zeros, each representing a localized topological charge. The number and nature of these charges depend sensitively on the system parameters, particularly the cloud of strings parameter $b$ and the coupling constant $\alpha$. For instance, when $b$ is small and $\alpha$ is varied, the system tends to develop two topological charges of opposite signs, resulting in a net total charge of zero. As $b$ increases, a transition occurs wherein the vector field develops three zeros, leading to a total topological charge of $+1$. This change in topological structure persists across different values of $\alpha$, underscoring the influence of these parameters on the BH’s thermodynamic classification. Our approach interprets the BH free energy as a scalar function defined over a two-dimensional parameter space—typically involving the horizon radius and an auxiliary angular variable. By constructing a vector field from the gradient of this free energy, we identify its zeros, which correspond to equilibrium states of the BH. Each zero carries a topological charge (winding number), determined by the rotation of vector field lines in its vicinity. These results naturally extend the classification framework of classical BHs. For example, Schwarzschild, Reissner-Nordström (RN), and AdS-Reissner-Nordström (AdS-RN) solutions correspond to total topological charges of $-1$, $0$, and $+1$, respectively. Our findings demonstrate that the parameters $b$ and $\alpha$ can continuously deform the topological structure of the BH solution space, thereby modulating the system’s thermodynamic phase behavior and stability profile.
Configurations with total charge $+1$ mimic the thermodynamics of AdS-RN BHs, while those with total charge $0$ resemble RN solutions without a cosmological constant. The structure and number of zeros in the vector field reflect the nature of possible phase transitions and the thermodynamic stability regime. The sensitivity of the topological charge to these parameters provides a powerful diagnostic for classifying BH phases under a wide range of physical conditions.

Finally, we investigated the dynamics of scalar field perturbations in the background of the chosen AdS BH (BH) solution coupled with a cloud of strings (CoS) and surrounded by a quintessence-like fluid (QF). To this end, we considered the massless Klein-Gordon equation for a minimally coupled scalar field propagating in the curved space-time geometry. By separating variables and applying a standard time-harmonic ansatz, the Klein-Gordon equation was reduced to a Schrödinger-like wave equation with an effective perturbative potential. This scalar potential encapsulates the influence of the geometric and physical parameters of the BH spacetime, including those associated with the CoS and QF. We derived the explicit form of this potential and examined how changes in the CoS parameter $\alpha$ and scale parameter $b$ affect its shape, height, and width. The resulting potential barrier governs the propagation and scattering behavior of scalar waves near the BH, and thus plays a crucial role in determining the characteristic quasinormal mode (QNM) spectrum.

To extract the QNM frequencies, we employed the WKB approximation method, which is well-suited for effective potentials with single-peaked barriers. Numerical computations were carried out for various values of the CoS parameter $\alpha$, while keeping the QF parameter and other BH parameters fixed. The analysis revealed that increasing $\alpha$ leads to noticeable shifts in both the real and imaginary parts of the QNM frequencies. Specifically, the real part-which corresponds to the oscillation frequency-increases, while the imaginary part-which governs the damping rate-generally decreases in magnitude, indicating longer-lived modes. 

These results highlight the significant impact of the CoS on the stability and dynamical response of the BH under scalar perturbations. In particular, they demonstrate how modifications to the background geometry alter the quasinormal spectra, which are in principle observable through gravitational wave signals. This analysis also provides further insights into the interplay between BH thermodynamics, geometric deformations, and field dynamics in curved spacetime.

\section*{Acknowledgments}

F.A. acknowledges the Inter University Centre for Astronomy and Astrophysics (IUCAA), Pune, India for granting visiting associateship.

\section*{Data Availability Statement}

This manuscript has no associated data.

\section*{Conflict of Interests}

Author declare (s) no conflict of interest.

\end{document}